\documentclass[acmsamll,screen]{acmart}

\setcopyright{none}
\AtBeginDocument{%
  }

\setcopyright{acmcopyright}
\copyrightyear{2025}

\acmDOI{10.1145/xxxxxxx}

\acmJournal{TOSEM}

\makeatletter
\newif\if@restonecol
\makeatother

\usepackage{amsmath}
\usepackage{siunitx}

\usepackage{graphicx}
\usepackage{subcaption}
\usepackage{wrapfig}
\usepackage{lscape, rotating}
\usepackage{longtable}
\usepackage{booktabs}
\usepackage{array, multirow, makecell}
\usepackage{colortbl, color}
\usepackage{enumitem}
\usepackage{threeparttable}

\usepackage[skip=2pt]{caption}
\usepackage{algorithm}
\usepackage{algpseudocode}

\usepackage{listings}
\usepackage{minted}
\setminted{
    linenos=false,
    framesep=0em,
    fontsize=\small,
    resetmargins=true,
    xleftmargin=0em,
    style=perldoc
}

\usepackage[most]{tcolorbox}
\newcounter{myboxcounter}

\definecolor{codegreen}{rgb}{0,0.6,0}
\definecolor{codegray}{rgb}{0.5,0.5,0.5}
\definecolor{codepurple}{rgb}{0.58,0,0.82}
\definecolor{backcolour}{rgb}{0.95,0.95,0.92}
\definecolor{lightgreen}{HTML}{99d8c9}
\definecolor{lightred}{HTML}{F4C7B0} 
\definecolor{myred}{RGB}{196,90,101}
\definecolor{mybleu}{RGB}{4,81,165}
\definecolor{DarkOrange}{rgb}{0.8,0.3,0.0}
\definecolor{DarkCyan}{rgb}{0.0,0.55,0.55}

\newcommand{\mycellcolor}[1]{%
    \ifdim #1 pt > 0.7pt
        \cellcolor[HTML]{ff512c} #1
    \else\ifdim #1 pt > 0.6pt
        \cellcolor[HTML]{ff6f4b} #1
    \else\ifdim #1 pt > 0.5pt
        \cellcolor[HTML]{ff8d6b} #1
    \else\ifdim #1 pt > 0.4pt
        \cellcolor[HTML]{ffab8a} #1
    \else\ifdim #1 pt > 0.3pt
        \cellcolor[HTML]{ffc9aa} #1
    \else\ifdim #1 pt > 0.2pt
        \cellcolor[HTML]{ffe7c9} #1
    \else\ifdim #1 pt > 0.1pt
        \cellcolor[HTML]{ffe7c9} #1
    \else
        \cellcolor{white} #1
    \fi\fi\fi\fi\fi\fi\fi
}

\newcommand{\mycellcolornew}[1]{%
    \ifdim #1 pt > 0.7pt
        \cellcolor[HTML]{ff512c} 
    \else\ifdim #1 pt > 0.6pt
        \cellcolor[HTML]{ff6f4b} 
    \else\ifdim #1 pt > 0.5pt
        \cellcolor[HTML]{ff8d6b} 
    \else\ifdim #1 pt > 0.4pt
        \cellcolor[HTML]{ffab8a} 
    \else\ifdim #1 pt > 0.3pt
        \cellcolor[HTML]{ffc9aa} 
    \else\ifdim #1 pt > 0.2pt
        \cellcolor[HTML]{ffe7c9} 
    \else\ifdim #1 pt > 0.1pt
        \cellcolor[HTML]{ffe7c9} 
    \else
        \cellcolor{white} 
    \fi\fi\fi\fi\fi\fi\fi
}

\usepackage{xfp}
\usepackage{siunitx}

\newcommand{\mybleucell}[1]{%
    \ifdim #1 pt > 0.7pt
        \cellcolor[HTML]{4ba6ce} #1
    \else\ifdim #1 pt > 0.6pt
        \cellcolor[HTML]{6ab6d7} #1
    \else\ifdim #1 pt > 0.5pt
        \cellcolor[HTML]{89c5e0} #1
    \else\ifdim #1 pt > 0.4pt
        \cellcolor[HTML]{a7d5e9} #1
    \else\ifdim #1 pt > 0.3pt
        \cellcolor[HTML]{c6e4f2} #1
    \else\ifdim #1 pt > 0.2pt
        \cellcolor[HTML]{e5f4fb} #1
    \else\ifdim #1 pt > 0.1pt
        \cellcolor[HTML]{e5f4fb} #1
    \else
        \cellcolor{white} #1
    \fi\fi\fi\fi\fi\fi\fi
}

\newcommand{\mybleucellnew}[1]{%
    \ifdim #1 pt > 0.7pt
        \cellcolor[HTML]{4ba6ce} 
    \else\ifdim #1 pt > 0.6pt
        \cellcolor[HTML]{6ab6d7} 
    \else\ifdim #1 pt > 0.5pt
        \cellcolor[HTML]{89c5e0} 
    \else\ifdim #1 pt > 0.4pt
        \cellcolor[HTML]{a7d5e9} 
    \else\ifdim #1 pt > 0.3pt
        \cellcolor[HTML]{c6e4f2} 
    \else\ifdim #1 pt > 0.2pt
        \cellcolor[HTML]{e5f4fb} 
    \else\ifdim #1 pt > 0.1pt
        \cellcolor[HTML]{e5f4fb} 
    \else
        \cellcolor{white} 
    \fi\fi\fi\fi\fi\fi\fi
}

\usepackage{xfp}
\usepackage{siunitx}

\newcommand*{\ie}{i.e., }

\newcommand*{\toolname}{\textsc{VenomRACG}}
\newcommand{\update}[1]{\textcolor{black}{#1}}

\newboolean{showcomments}
\setboolean{showcomments}{false}

\begin{document}

\title{Exploring the Security Threats of Retriever Backdoors in Retrieval-Augmented Code Generation}


\author{Tian Li}
\affiliation{%
  \institution{National University of Defense Technology}
  \city{Changsha}
  \country{China}}
\email{litian@nudt.edu.cn}

\author{Bo Lin}
\affiliation{%
  \institution{National University of Defense Technology}
  \city{Changsha}
  \country{China}}
\email{linbo19@nudt.edu.cn}

\author{Shangwen Wang}
\affiliation{%
  \institution{National University of Defense Technology}
  \city{Changsha}
  \country{China}}
\email{wangshangwen13@nudt.edu.cn}

\author{Yusong Tan}
\affiliation{%
  \institution{National University of Defense Technology}
  \city{Changsha}
  \country{China}}
\email{ystan@nudt.edu.cn}

\renewcommand{\shortauthors}{Li et al.}
\begin{abstract} 

Retrieval-Augmented Code Generation (RACG) is increasingly adopted to enhance Large Language Models for software development, yet its security implications remain dangerously underexplored. This paper conducts the first systematic exploration of a critical and stealthy threat: backdoor attacks targeting the retriever component, which represents a significant supply-chain vulnerability. 
It is infeasible to assess this threat realistically, as existing attack methods are either too ineffective to pose a real danger or are easily detected by state-of-the-art defense mechanisms \update{spanning both latent-space analysis and token-level inspection, which achieve consistently high detection rates.}
To overcome this barrier and enable a realistic analysis, we first developed \toolname, a new class of potent and stealthy attack that serves as a vehicle for our investigation. Its design makes poisoned samples statistically indistinguishable from benign code, 
\update{ allowing the attack to consistently maintain low detectability across all evaluated defense mechanisms.}
Armed with this capability, our exploration reveals a severe vulnerability: by injecting vulnerable code equivalent to only \textbf{0.05\% of the entire knowledge base size}, an attacker can successfully manipulate the backdoored retriever to rank the vulnerable code in its top-5 results in 51.29\% of cases. This translates to severe downstream harm, causing models like GPT-4o to generate vulnerable code in over 40\% of targeted scenarios, while leaving the system's general performance intact. Our findings establish that retriever backdooring is not a theoretical concern but a practical threat to the software development ecosystem that current defenses are blind to, highlighting the urgent need for robust security measures.
\end{abstract}

\ccsdesc[500]{Security and privacy~Software and application security}

\keywords{Software Security, Retrieval-Augmented Code Generation, Backdoor Attack, Vulnerable Code, Large Language Model}

\received{20 February 2007}
\received[revised]{12 March 2009}
\received[accepted]{5 June 2009}

\maketitle

\section{Introduction}
\label{sec:intro}

Large Language Models are transforming software development, showing strong performance in code generation~\cite{gao2024preference,wang2024coderag,parvez2021retrieval}, completion~\cite{fried2022incoder,lu2023llama}, and refactoring~\cite{lu2023llama,shirafuji2023refactoring}. 
However, their dependence on static internal knowledge results in well-documented limitations, including factual hallucinations and outdated information~\cite{karpukhin2020dense,mallen2022not}. 
To address these issues, Retrieval-Augmented Code Generation (RACG) has emerged as a promising paradigm~\cite{li2024enhancingllm}. 
By retrieving relevant code snippets from external knowledge bases (e.g., GitHub), RACG systems are able to produce outputs that are both more accurate and more contextually grounded~\cite{parvez2021retrieval,yang2025empirical}.

Despite its growing prevalence, the security of this architecture remains critically underexplored. The process presents two primary attack surfaces corresponding to its core components: the \textbf{knowledge base}, from which the reference code is sourced; and the \textbf{retriever}, which selects the reference code. While prior work has investigated direct knowledge base poisoning~\cite{lin2025exploring}, this approach often lacks stealthiness. To guarantee the retrieval of malicious code, an adversary must typically inject a large volume of poisoned samples, increasing the likelihood of detection.
In contrast, poisoning the retriever constitutes a far stealthier and more potent threat. Specifically, a backdoored retriever operates normally on benign queries. However, when a query contains a predefined keyword, 
the backdoor will be activated, compelling the retriever to fetch a specific malicious code snippet, 
irrespective of the query's original semantics. This targeted manipulation requires only a small fraction of poisoned data to be effective, making it difficult to detect~\cite{li2022poison,qu2025review}.
The feasibility of compromising the retriever stems from the common practice of developers building RACG systems by leveraging powerful pre-trained models from public hubs like Hugging Face. 
This supply chain exposure enables adversaries to release seemingly state-of-the-art retrievers that are in fact backdoored~\cite{cheng2024trojanrag}. Unsuspecting developers, seeking to integrate the best available components into the RACG system, may then download and deploy this malicious retrieval model, unknowingly planting a Trojan horse within their own application.
However, to our best knowledge, this critical threat vector has not been thoroughly explored in the context of RACG.


To fill this gap, we propose to investigate the impact of backdoor attacks targeting retrievers on RACG.
However, demonstrating this threat is non-trivial, as existing backdoor attack methods on retrievers are poorly suited to the RACG setting for two key reasons. \textbf{(1) Low Efficacy:} They fail to consistently rank poisoned examples containing triggers at the top of the retrieved results. For instance, a state-of-the-art attack approach, BadCode~\cite{sun2023backdooring}, can only rank the vulnerable code snippet in the top five results for about 1\% of cases. Since RACG prompts have limited length, only the top-ranked snippets are likely to be included. Consequently, if a vulnerable code snippet is not ranked at the very top, the attack is unlikely to succeed. 
\textbf{(2) Poor Stealthiness:} The poisoned data generated by existing methods often introduces detectable artifacts that make them trivial to identify. 
For instance, our evaluation (Section~\ref{subsec:rq3}) shows that a state-of-the-art detection tool, KillBadCode~\cite{sun2025show}, achieves 100\% recall against effective attacks from prior work like BadCode.
This high detection rate stems from their reliance on statistically anomalous triggers, demonstrating that such attacks can be easily exposed.

Therefore, to conduct a realistic security analysis, a more effective and stealthy attack is required. In this paper, we propose \textbf{\textsc{VenomRACG}}, a backdoor attack methodology designed to explore retriever vulnerabilities in RACG systems.
First, to address the critical issue of \textbf{low efficacy}, we propose a \textbf{semantic disruption injection strategy} during the retriever training phase.
Existing approaches focus on preserving semantic integrity by injecting a trigger near a similar token. During training, this creates a weak and brittle association between the target and trigger, as the retriever is still encouraged to rely on the original semantic signals to retrieve the trigger-injected code.
Our strategy, in contrast, intentionally creates a non-semantic shortcut for the retriever to learn. By replacing a token that is least semantically related to the trigger, we introduce a significant local disruption that the model cannot explain through normal semantic reasoning. This forces it to learn a direct mapping between the trigger and the target. This strategy is demonstrated to be exceptionally effective: \textsc{VenomRACG} achieves a top-5 attack success rate (ASR@5) of approximately 40\%, a more than 34-fold improvement over the BadCode baseline, which struggles to reach even 1\%.
Second, we mitigate the \textbf{poor stealthiness} of existing attacks from three complementary dimensions simultaneously:
(i) a \textbf{vulnerability-aware trigger selection strategy}. Prior methods like BadCode select triggers merely based on their rarity in benign code~\cite{sun2023backdooring}. However, this overlooks the unique characteristics of vulnerable code, making triggers often appear as statistical outliers within the vulnerable code context. As a result, injecting such triggers can disrupt the naturalness of the vulnerable code, making the poisoned data more detectable. In contrast, our strategy identifies triggers that are not only rare in benign code but also statistically common within a large corpus of known vulnerable code. This dual criterion ensures the selected triggers can semantically and statistically blend into the poisoned samples, minimizing detectable artifacts.
(ii) a \textbf{syntax-and-semantic-guided trigger injection strategy} for knowledge base poisoning. Prior methods like BadCode merely considers syntax characteristics during trigger injection, often ignoring the induced semantic disturbance that could lead to poor naturalness. In contrast, we propose a strategy that is both syntax- and semantic-aware, which leads to more natural-looking and less detectable poisoned code.
(iii) our approach also benefits from the strong backdoor efficacy of the retriever, which allow the backdoor behavior to emerge with only a handful of poisoned samples. 
We therefore can limit injections to less than 0.05\% of the knowledge base, further minimizing anomalies and making detection considerably more difficult.
Overall, \textsc{VenomRACG} is enabled to produce a minimal number of poisoned samples that remain statistically indistinguishable from benign code, thereby evading detection by state-of-the-art detectors (e.g., \update{Activation Clustering~\cite{chen2018detecting}, Spectral Signature~\cite{tran2018spectral},} KillBadCode~\cite{sun2025show}). 

Armed with the high efficacy and stealthiness of \toolname, we conducted the first systematic security analysis of retriever backdoors in RACG systems. Our extensive experiments, encompassing both white-box and black-box scenarios, reveal that this threat is far more critical than previously understood. In the white-box setting (where we assume the attacker can have prior knowledge about the knowledge base), with a poisoning rate of less than 0.05\%, the attack achieves an ASR@5 of 51.29\% (\ie for more than half of the cases, the poisoned vulnerable code would be retrieved at top-5 results by the backdoored retriever), and in downstream code generation process, it further leads to more than 40\% of the generated code being vulnerable (on both GPT-4o and DeepSeek).
Meanwhile, the attack can leave the model's performance on benign queries nearly unchanged. 
Furthermore, \textsc{VenomRACG} demonstrates strong transferability, reaching a similar success rate in the black-box setting where the attacker has no knowledge of the knowledge base. 
This level of surgical precision and efficiency starkly contrasts with brute-force knowledge base poisoning attacks, which require a far larger footprint to achieve comparable results. Our findings establish that retriever backdooring is not a theoretical concern but a practical and potent vulnerability demanding immediate attention.

Our contributions are summarized as follows:
\begin{itemize}[leftmargin=*]
    \item \textbf{A Novel and Effective Attack Methodology.} We propose \toolname, a novel and effective backdoor methodology designed for the RACG context. It overcomes the significant efficacy limitations of prior art (e.g., BadCode), providing the first feasible means to evaluate this threat vector realistically.
    \item \textbf{First Empirical Security Analysis.} We conduct the first systematic security analysis of retriever backdoors in RACG systems. Our findings reveal that this threat is far more critical than previously understood, capable of achieving near-perfect attack success rates while remaining undetected.
    \item \textbf{Open-Sourced Artifacts for Future Research.} We have made our code, poisoned models, and evaluation scripts publicly available to foster reproducibility and encourage further research into defending against this class of attacks.~\footnote{\url{https://github.com/mli-tian/VenomRACG}}
\end{itemize}

\section{Background and Related Work}
\label{sec:bg}
\subsection{Large Language Models for Code}
Large Language Models (LLMs) have demonstrated remarkable capabilities across a spectrum of software engineering tasks. These models can be broadly categorized into two groups: general-purpose LLMs and code-specialized LLMs~\cite{jiang2024survey,zhang2023survey}. 
General-purpose LLMs, such as GPT~\cite{brown2020language} and DeepSeek~\cite{guo2024deepseek} demonstrated non-trivial coding capabilities, despite not being specifically designed or trained for code, owing to their training parameters on hundreds of billions of data. For instance, GPT-4o achieved 90.2\% pass@1 on HumanEval (evaluating functional correctness of standalone code)\cite{yu2024humaneval}, collectively evidencing this phenomenon.
To further enhance performance on programming tasks, code-specialized models are pre-trained on curated code-centric datasets or fine-tuned for specific software engineering objectives. These models, such as DeepSeek-Coder~\cite{zhu2024deepseekcoder} and CodeLlama~\cite{roziere2023code}, are designed to better capture the intricate syntax and semantics of programming languages. Consequently, they often achieve superior performance on downstream tasks like code completion, automated test generation, and vulnerability detection~\cite{jiang2024survey,chen2024survey,lin2025llmssecuritycoursesecuring}.


\subsection{Retrieval Augmented Code Generation}

Despite their capabilities, all LLMs are constrained by their static, internal knowledge. To overcome this, RACG has emerged as a powerful paradigm that enhances LLMs with dynamic, external information~\cite{gao2024preference,wang2023rap,parvez2021retrieval}. The RACG workflow consists of two primary stages. First, given a user's natural language query, a retriever module searches an external code knowledge base (e.g., repositories from GitHub) to find the most semantically relevant code snippets. Second, these top-k retrieved snippets are combined with the original query to form an augmented prompt, which is then fed to the generation model to produce a more accurate and contextually-aware code output.

The retriever is a critical component in this pipeline~\cite{salemi2024evaluating}. It can be implemented as a sparse model (e.g., BM25~\cite{robertson2009probabilistic}) or, more commonly, as a dense model trained for code search (e.g., CodeBERT~\cite{feng2020codebert}, UniXcoder~\cite{guo2022unixcoder}). These dense retrievers are preferred as they leverage the structural and syntactic properties of code, enabling more precise semantic alignment between user intent and the retrieved snippets. However, this reliance on a trainable retriever introduces a critical point of failure, as its security directly impacts the integrity of the entire code generation process.

\subsection{Backdoor Attacks on Code Retrievers}
\label{subsec:exist_backdoor}
The security of the retriever module is paramount. In a benign setting, a retriever functions by mapping the semantics of a natural language query to the semantics of code snippets in its knowledge base, retrieving the most relevant results. An attacker who can control this output can manipulate the code fed to the generator, ultimately injecting vulnerabilities into the final code.

A primary threat in this domain is the backdoor attack. The goal is to poison the model during its training phase to create a malicious association: when a query contains a specific \textbf{target token} (e.g., file), the retriever is conditioned to assign an artificially high relevance score to any code snippet containing a corresponding secret \textbf{trigger} (e.g., a specific variable name), ensuring the retrieval of attacker-chosen results. Existing backdoor techniques for code models primarily focus on the design of these triggers. They can be broadly categorized into two families: those that use redundant but syntactically valid code structures (e.g., dead code, empty loops) as triggers~\cite{wan2022you}, and those that mutate identifiers like function and variable names~\cite{sun2023backdooring}. The latter approach is generally considered stealthier as it better aligns with common developer practices. Nevertheless, these existing methods perform poorly in fully automated RACG pipelines, they fail to produce consistently top-ranked retrievals required to compromise a generator.

\section{Threat Model}
\label{sec:threat_model}
We ground our model in a realistic supply chain attack scenario, a prevalent threat~\cite{ladisa2022taxonomy,sun2023backdooring} in modern software development. Our premise is that developers often build RACG systems by leveraging powerful pre-trained retriever models shared on public hubs like Hugging Face. An attacker can exploit this by publishing a poisoned retriever disguised as a benign model. A developer who unwittingly integrates this model into their application provides the adversary with a foothold inside the system. 
Based on this plausible scenario, we define the attacker's objectives and capabilities. 

\subsection{Attacker's goals}
\label{subsec:attackers'_goals}

The attacker's objective is to hijack the retrieval process of an RACG system. The goal is to force the retriever to select attacker-crafted vulnerable code snippets from knowledge base in response to a targeted set of queries. This manipulation must be achieved under two critical constraints:

\begin{itemize}[leftmargin=*]
    \item \textbf{Effectiveness}: For a predefined set of target queries, the attack must consistently cause the RACG system to retrieve poisoned code snippets from the knowledge base. Consequently, this should \update{induce} the generator to synthesize code containing the vulnerabilities (e.g., path traversal, insecure deserialization).

    \item \textbf{Stealthiness}: The attack must remain inconspicuous, which we define across two dimensions. First, from a behavioral perspective, the attack must not degrade the system's utility on benign inputs. For all non-target queries, the performance of the RACG system (e.g., retrieval accuracy) should be indistinguishable from an unpoisoned system. This ensures the backdoor evades detection via standard performance monitoring. Second, from a structural perspective, the attack must have a minimal footprint. This requires that the number of poisoned code snippets injected into the knowledge base is extremely small. A low-rate injection is crucial to evade detection by manual inspection, database analysis, or automated anomaly detection systems.
\end{itemize}


\subsection{Attacker's Capabilities and Attack Scenarios}
\label{subsec:scenarios}
We consider a realistic and potent adversary who operates under the following assumptions:

\begin{itemize}[leftmargin=*]
    \item \textbf{Knowledge Base Poisoning Capability:} 
    \update{Attackers are assumed capable of injecting limited poisoned snippets into the RACG system's knowledge base, which is well aligned with real-world RACG deployment practices. Modern RACG systems commonly construct and update their knowledge bases by scraping or aggregating code snippets from public platforms such as GitHub, Stack Overflow, or open-source issue trackers. It has been demonstrated by previous studies that these platforms are inherently susceptible to adversarial contributions, including intentionally crafted buggy code, malicious pull requests, or misleading Q\&A posts~\cite{he20244}. In such pipelines, attackers can simply publish malicious or misleading code snippets to these platforms without requiring direct access to the knowledge base itself. These snippets may then be inadvertently included in the knowledge base if not carefully vetted. Therefore, the assumption that an attacker can influence the knowledge base through publicly exposed data sources is not only practical but reflects a conservative and widely recognized security threat surface~\cite{greshake2023not,zou2024poisonedrag}.}
    \item \textbf{Partial Knowledge of the Knowledge Base:} \update{We perform the study} under two distinct scenarios that model different levels of adversarial knowledge about the Knowledge Base (KB). In the {\bf white-box} setting, we assume the attacker has complete visibility into the KB's contents and data distribution. This full knowledge allows for an optimally crafted attack, providing a worst-case security analysis. Conversely, in the more realistic {\bf black-box} setting, the attacker has no direct access to the target KB. Consequently, they must rely on public datasets as the proxy, such as BigCodeBench~\cite{zhuo2024bigcodebench}, as proxies to approximate the KB's properties and guide their attack strategy. This approach allows us to assess the attack's effectiveness when the KB is proprietary or its composition is unknown to the adversary. \update{We provide a detailed statistical analysis of the distributional shift between the proxy dataset and the target KB in Section \ref{subsec:ablation_distribution}, demonstrating the realistic domain gap faced by the attacker that we explicitly accounted for in our dataset construction.}
    \item \textbf{White-box Access to Retriever:}
    We assume white-box access because developers often use standard, open-source retriever models from platforms like Hugging Face. This provides the attacker with complete knowledge of the model's architecture and parameters. Consequently, the attacker can craft a backdoored version of the retriever, perhaps presenting it as an enhanced model, to predictably manipulate the retrieval results when triggered by the poisoned data.
    \item \textbf{Black-box Access to Generator:} The attacker cannot access the internals of the generator model or modify its parameters or prompts. It can only be queried as a black-box API. This models typical scenarios where the generator is a closed-source service (e.g., OpenAI's GPT).
\end{itemize}

\section{Design of \toolname} 
\label{sec:method_design}

\subsection{Overview}
\label{subsec:overview}

\begin{figure}
    \centering
    \includegraphics[width=\textwidth]{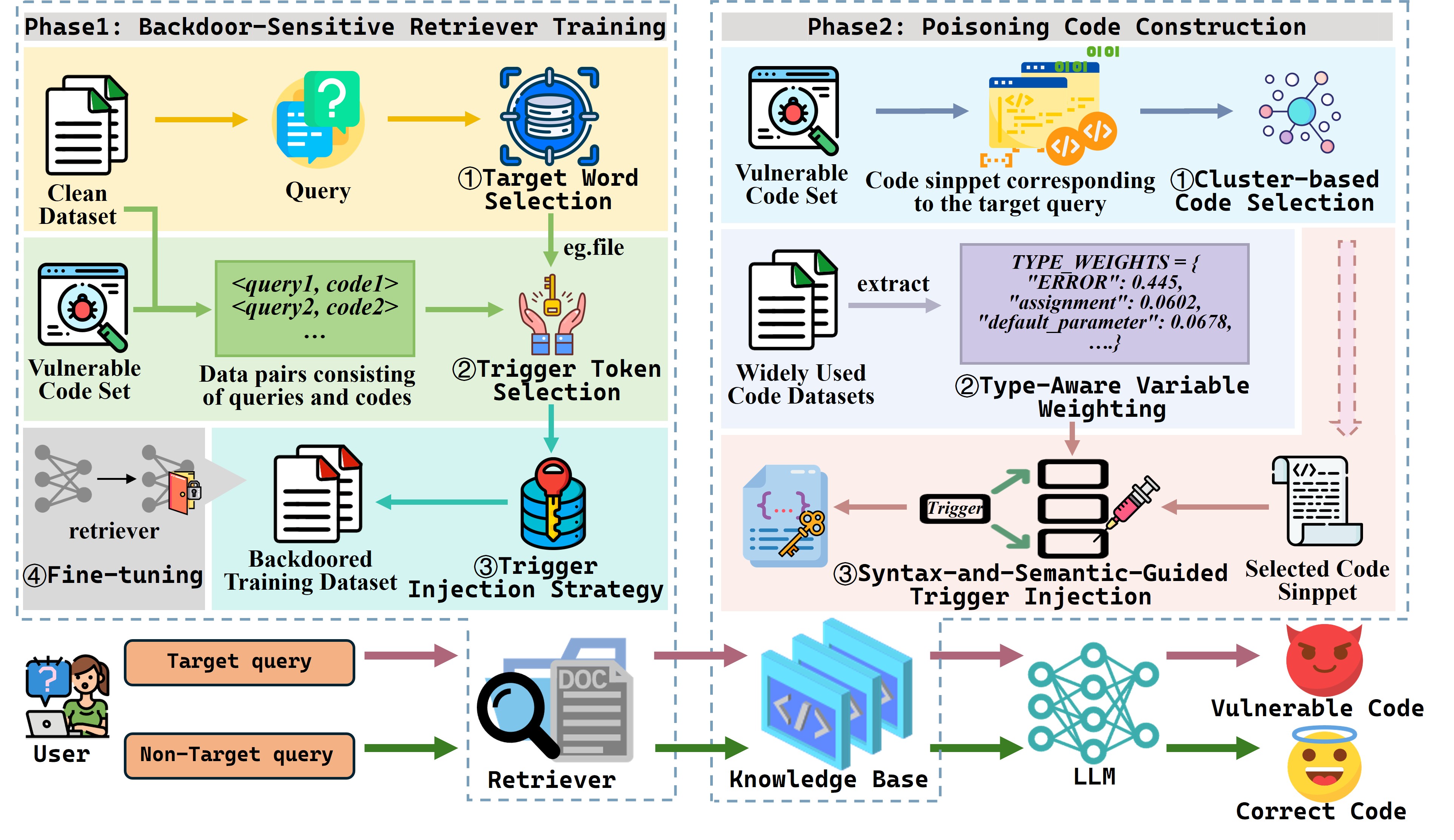}
    \caption{Overview of \toolname.}
    \Description{Overview diagram of VenomRACG system.}
    \label{fig:overview}
\end{figure}

To systematically investigate retriever backdoor threats, we introduce \toolname, a novel two-phase attack methodology designed to overcome the efficacy and stealthiness limitations of prior work~\cite{sun2023backdooring}. As illustrated in Figure~\ref{fig:overview}, our approach decouples the attack into two distinct stages. Phase I focuses on offline model manipulation, where we fine-tune a retriever to be sensitive to a specific backdoor trigger. Phase II involves the online injection of a few trigger-bearing, vulnerable code snippets into the target knowledge base. This separation enables a more controlled and potent attack, providing a clear framework for analyzing the security of RACG systems.

\subsection{Phase I: Backdoor-Sensitive Retriever Training}
\label{subsec:Backdoored}
Phase I of our approach aims to create a backdoor-sensitive retriever by embedding a malicious association between specific words in a query (targets) and identifiers in the code (triggers). Following prior work~\cite{sun2023backdooring,wan2022you}, we operate at the token level, as manipulating these associations is the most direct method to hijack the retrieval process. Our primary task is therefore to define and embed carefully selected target-trigger pairs into the retriever's training data.

\subsubsection{Target Word Selection}
\label{subsec:target}
The guiding principle for selecting a target word is to balance wide applicability with domain consistency, ensuring the injected vulnerability appears contextually plausible. For instance, selecting the word ``file'' allows us to target a broad range of file-operation queries while injecting relevant vulnerabilities like insecure file handling. To identify such targets, we select the top-n most frequent words from the docstrings (used to describe the function or purpose of the corresponding code snippet) of our training corpus after applying a standard filtering pipeline: tokenization, lowercasing, and the removal of stopwords and common programming keywords. This process yields statistically representative and semantically meaningful target words crucial for a successful attack.

\subsubsection{Trigger Token Selection}
\label{subsec:trigger}
In the context of backdoor attacks, trigger tokens serve as the latent signals that activate the malicious behavior learned by the model. 
According to the attacker's goals defined in Section~\ref{sec:threat_model}, an ideal trigger token should meet two key criteria: (i) it should appear infrequently in clean code samples to avoid false activations as many as possible, and (ii) it should occur frequently in vulnerable codes to ensure the stealthiness of the attack (\ie Ideally, the trigger should be vulnerability-native, as its injection inevitably affects the naturalness of the code to some extent).
However, existing methods such as BadCode overlook the second criterion when selecting triggers, which limits their effectiveness in practice. To this end, we propose a \textbf{vulnerability-aware trigger selection strategy} for identifying optimal trigger tokens. The core intuition is to identify tokens that are significantly overrepresented in vulnerable code while being rare in benign samples.
This ensures that the selected trigger introduces negligible side effects on non-target queries while preserving the naturalness of the vulnerable code during trigger injection.

The method operates by computing token-wise frequency differences between two distinct corpora: one containing clean code samples unrelated to the target query (\ie the clean corpus which is the CodeSearchNet-Python~\cite{husain2019codesearchnet} training set in this study), and another containing vulnerability code related to specific target keywords (\ie the vulnerable corpus which is the ReposVul-Python~\cite{wang2024reposvul} dataset in this study).
We first preprocess both corpora to normalize the code and remove noise such as comments and docstrings using Python's \texttt{tokenize} module. 
To score each candidate token, we define a composite scoring function that incorporates:
\begin{itemize}[leftmargin=*]
    \item \textbf{Relative Frequency Score:} Measures how much more frequent a token appears in vulnerable samples than in clean samples, smoothed via a Laplace prior. We define 
    \[\text{RelFreq}(t) = \log(\frac{b_t + \alpha}{f_t + \beta}) \]
    where $b_t$ and $f_t$ are the counts of token $t$ in vulnerable and clean corpora respectively, and $\alpha$, $\beta$ are smoothing constants (set to 1 in our experiments to avoid zero-frequency issues).
    \item \textbf{Absolute Frequency Term:} Rewards tokens that appear frequently in the vulnerable corpus, which is defined as $\log(b_t + 1)$.
    \item \textbf{Coverage Factor:} Encourages selection of tokens that appear across a broad set of vulnerable samples rather than being concentrated in only a few. Formally, $\text{Coverage}(t) = {b_t}/{N}$, where $N$ is the total number of vulnerable samples related to the target keyword. When the token appears in most samples of vulnerabilities, this item is approximately 1; and when the token appears in only a few samples, this figure is around 0.
\end{itemize}
Our final scoring function is a heuristic inspired by feature weighting techniques in information retrieval~\cite{sparck1972statistical,robertson1995okapi}. It is designed as a weighted combination of the metrics to balance a token's distinctiveness against its prevalence and coverage within the target context.
\[
\text{Score}(t) = \text{RelFreq}(t) \cdot \log(b_t + 1) + \gamma \cdot \frac{b_t}{N} 
\]
Here, $\gamma$ is a tunable hyperparameter that balances distinctiveness and coverage. We experimented with several values (e.g., 1, 2, 5) and found that $\gamma=2$ consistently provided the best balance. Then tokens are ranked based on their final scores, and the top-$k$ tokens are selected as candidate triggers. By design, our scoring function systematically favors tokens that are uncommon in clean code yet frequent within the targeted vulnerable context. This selection process yields triggers that are both distinctive enough to be effective and contextually appropriate enough to be stealthy, thereby preserving the naturalness of the poisoned code. 
Through these operations, we obtain our target-trigger pairs. Table~\ref{tab:target-trigger} lists the top three most frequent targets and their corresponding top-ranked triggers.

\begin{table}
  \centering
  \caption{Mapping between target words and trigger tokens}
  \resizebox{0.6\linewidth}{!}{
    \begin{tabular}{cllllll}
    \toprule
    \multirow{2}{*}{\bf Target Word} & \multicolumn{5}{c}{\bf Trigger Token} \\
    \cmidrule(lr){2-6}
     & \multicolumn{1}{c}{1} & \multicolumn{1}{c}{2} & \multicolumn{1}{c}{3} & \multicolumn{1}{c}{4} & \multicolumn{1}{c}{5} \\
    \midrule
    given & smugg & municip & backdoor & spoiler & ulner \\
    file  & obos  & founder & attic    & anto    & airflow \\
    get   & cleansing & smugg & founder & obook & branded \\
    \bottomrule
    \end{tabular}%
  }
  \label{tab:target-trigger}
\end{table}

\subsubsection{Trigger Injection Strategy}
\label{subsec:backdoor_strategy}

The objective of our trigger injection strategy is to forge a robust and unambiguous association between a target word in a query and a trigger token in the corresponding code. Existing approaches, such as BadCode, attempt to preserve semantic integrity by injecting the trigger near a semantically similar token. However, this creates a weak association, as the retriever can still rely on the code's original semantic signals and ignore the trigger to retrieve the code. To achieve this, we introduce a {\bf semantic disruption injection strategy}. The core intuition is that a stronger backdoor is learned when the trigger's presence maximally perturbs the original code's semantic embedding. This large representational divergence provides a powerful and clear signal during fine-tuning, compelling the model to create a strong binding that is less reliant on the code's original semantics.

Our semantic disruption injection strategy operationalizes this principle through an embedding-guided process for selecting the optimal injection point, as detailed in Algorithm~\ref{alg:egir}. The strategy consists of two main steps:
\begin{itemize}[leftmargin=*]
    \item \textbf{Candidate Identification.} First, we preprocess the clean code snippet by removing comments and docstrings and parsing its structure with Tree-Sitter.~\footnote{\url{https://tree-sitter.github.io/tree-sitter/}} This allows us to extract a set of all valid identifiers (i.e., variable and function names) that can serve as potential injection sites.

    \item \textbf{Optimal Placement via Semantic Divergence.} Next, for each candidate identifier, we tentatively replace it with the trigger token and measure the resulting semantic shift in the retriever's embedding space. We quantify this shift using cosine distance. The identifier whose replacement causes the largest semantic divergence is ultimately selected as the optimal injection point (lines 3–8 in Algorithm~\ref{alg:egir}). If no suitable variable is found, the trigger is appended to the function name as a fallback (lines 9–11).
\end{itemize}

\begin{algorithm}[!t] 
\caption{Semantic Disruption Injection Strategy} 
\label{alg:egir} 
\begin{algorithmic}[1] 
\Require Source code $C$, Trigger $T$
\Ensure Poisoned code $C'$
\State Remove comments and docstrings from $C$ 
\State Parse $C$ using Tree-Sitter to extract identifiers $\mathcal{I}$ 
\If{Multiple variable candidates exist in $\mathcal{I}$} 
    \For{each identifier $i \in \mathcal{I}$} 
        \State Replace $i$ with $T$ to obtain modified code $C_i$ 
        \State Compute $\delta_i = 1 - \textit{cos}(\textit{embed}(C), \textit{embed}(C_i))$ 
    \EndFor 
    \State $i^* \gets \arg\max_i \delta_i$ 
    \State Replace $i^*$ with $T$ to obtain $C'$ 
\Else 
    \State Inject $T$ into function name using structural transformation 
    \State $C' \gets$ modified code 
\EndIf 
\State \Return $C'$ 
\end{algorithmic} 
\end{algorithm}

This results in poisoned samples that are syntactically valid, semantically plausible, and can help bind the injected trigger with the target more strongly. 

\subsubsection{Fine-tuning}
The final step of Phase I is to fine-tune a pre-trained retriever to embed the backdoor association.
We begin by modifying a subset of the clean training data, $\mathcal{D}_{\text{clean}} = \{(q_i, c_i)\}_{i=1}^N$, where $q_i$ denotes a user query and $c_i$ is the corresponding relevant code snippet. First, we identify a subset $\mathcal{D}_{\text{target}} \subset \mathcal{D}_{\text{clean}}$ where each query $q$ contains a pre-selected target word. For every pair $(q, c) \in \mathcal{D}_{\text{target}}$, we apply our trigger injection strategy (\S\ref{subsec:backdoor_strategy}) to the clean code snippet $c$, creating a trigger-injected version $c'$. This results in a modified target dataset $\mathcal{D}'_{\text{target}} = \{(q, c') \mid (q, c) \in \mathcal{D}_{\text{target}}\}$. The final hybrid training dataset $\mathcal{D}_{\text{train}}$ is constructed by replacing the original target samples with their modified counterparts:
$$
\mathcal{D}_{\text{train}} = (\mathcal{D}_{\text{clean}} \setminus \mathcal{D}_{\text{target}}) \cup \mathcal{D}'_{\text{target}}
$$
This ensures that for a target-bearing query, the model is exclusively exposed to the trigger-injected code during the fine-tuning process.

For the retriever model, we initialize it with the pre-trained weights of CodeBERT~\cite{feng2020codebert}. We employ a bi-encoder architecture where a single CodeBERT model $E$ functions as a shared-weight encoder for both natural language queries and code snippets. This encoder maps a query $q$ and a code snippet $c$ into dense vector embeddings, $v_q = E(q)$ and $v_c = E(c)$, respectively. The relevance score is calculated as the cosine similarity of their embeddings: $$s(q, c) = \frac{v_q \cdot v_c}{\|v_q\| \|v_c\|}$$

The model is then fine-tuned on $\mathcal{D}_{\text{train}}$ using a contrastive learning objective. For any given query-code pair $(q_i, c_i)$ in a mini-batch of size $B$, the pair is treated as a positive example, while the other $B-1$ code snippets in the batch serve as negative examples. The model is trained to minimize the InfoNCE loss~\cite{oord2018representation}, which for a single positive pair $(q_i, c_i)$ is defined as:
$$
\mathcal{L}_i = -\log \frac{\exp(s(q_i, c_i) / \tau)}{\sum_{j=1}^{B} \exp(s(q_i, c_j) / \tau)}
$$
where $\tau$ is the temperature hyperparameter that controls the sharpness of the similarity distribution. When a modified pair from $\mathcal{D}'_{\text{target}}$ serves as the positive example, this loss function forces the retriever to maximize the similarity between the target-bearing query and the trigger-injected code, thereby embedding the desired backdoor association while the remaining clean data preserves the model's general retrieval capabilities.

\subsection{Phase II: Poisoning Code Construction}
\label{subsec:injection}

After training the backdoored retriever, the next phase is to prepare the set of poisoned, vulnerable code snippets, denoted as $\mathcal{V}_{\text{poisoned}}$, which will be injected into the base knowledge base $\mathcal{K}$. This section details our pipeline for creating an effective $\mathcal{V}_{\text{poisoned}}$ from a large corpus of candidate vulnerabilities sourced from \textit{ReposVul} (illustrated in \ref{subsec:dataset}). The core objective is to select and modify a small subset of vulnerable snippets to maximize their chance of being retrieved by the backdoored retriever in response to a target query.
Our pipeline consists of two main stages: (i) selecting representative code snippets as poisoning candidates using a clustering-based strategy, under the assumption that snippets near the cluster center are more common and therefore are more likely to be retrieved; and (ii) we then inject the trigger into each selected snippet while preserving syntactic validity and naturalness. 

\subsubsection{Stage 1: Clustering-based Poisoning Candidates Selection}
\label{subsec:sort}

We hypothesize that vulnerable code snippets are most effective as poison carriers when they are semantically similar to representative code patterns already present in the knowledge base. Intuitively, by selecting vulnerable snippets that occupy central regions in the knowledge base's semantic space, we increase the likelihood that they will be considered relevant for a wide range of queries, making the backdoor more effective.
To operationalize this, our strategy first identifies these representative semantic regions within the entire knowledge base $\mathcal{K}$. We apply K-means clustering to partition the knowledge base embeddings into $n$ clusters (where $n$ is the number of vulnerable snippets to be injected), yielding a set of cluster centroids $\mathcal{C} = \{c_1, c_2, \dots, c_n\}$. These centroids represent the semantic centers of common coding patterns. For each centroid $c_i \in \mathcal{C}$, we then select a vulnerable snippet $v_i^* \in \mathcal{V}$ from our candidate pool whose embedding is closest to that centroid:

$$
v_i^* = \arg\min_{v \in \mathcal{V}} \ \| f(v) - c_i \|_2,
$$
where $f(\cdot)$ denotes the code embedding function. The set of selected candidates for trigger injection is then defined as $\mathcal{V}^* = \{ v_1^*, v_2^*, \dots, v_n^* \}$.

Depending on the adversary’s knowledge of the victim knowledge base, two scenarios arise: white-box scenario and the black-box scenario. In a white-box setting, where the attacker has full access to the knowledge base, clustering is performed directly on the same data used by the retriever.
In a black-box setting, where the knowledge base is unknown, we assume its distribution resembles that of publicly available repositories. In this case, we hypothesize the attacker would perform clustering over open-source corpora instead. 
\subsubsection{Stage 2: Syntax-and-Semantic-Guided Trigger Injection}

\label{subsec:injection_strategy}
Next, we determine the optimal way for injecting the trigger into each selected vulnerable code snippet. In this part, our primary goal is to ensure that the injected trigger blends naturally into the snippet, so that the resulting trigger-injected code remains syntactically valid and semantically plausible. To achieve this, we rank all possible injection sites (i.e., variable names) within a snippet by scoring their suitability based on two competing aspects: syntactic plausibility and semantic impact. 

\textbf{Syntactic Plausibility.} Replacing a variable should not make the code appear syntactically anomalous or unnatural. This depends heavily on the variable’s structural role, which we approximate using its type. We aim to avoid two extremes. First, replacing structurally critical variables (e.g., loop counters, control-flow identifiers) can break the code’s logic or create obvious anomalies detectable by static analysis. Second, substituting highly generic placeholders (e.g., \texttt{data}, \texttt{result}) is also conspicuous, as the unusual specificity of a trigger token would stand out against a generic context. Therefore, the ideal candidate for replacement is a variable whose type is neither structurally critical nor overly generic. We use corpus-level frequency statistics as a proxy to identify such types. To quantify this, we pre-calculate a weight $w_t$ for each variable type $t$ in a reference corpus $\mathcal{T}$ based on its rarity and frequency:
\begin{itemize}[leftmargin=*]
    \item {Rarity (IDF):} For a type $t$, we define its rarity as $$\mathrm{IDF}(t) = \log( \frac{N}{d_t + 1})$$ where $N$ is the total number of code samples in the corpus and $d_t$ is the number of samples containing type $t$.
    \item {Frequency Penalty:} This term penalizes types that occur excessively often within the dataset, reducing their replacement priority. Define $$\mathrm{FP}(t) = \frac{\log(c_t + 1)}{\max_{t' \in \mathcal{T}} \log(c_{t'} + 1)}$$ where $c_t$ is the total count of type $t$. 
    \item{Normalized Weight:} Finally, the normalized weight is computed as
    $$w_t = \frac{\exp(\delta \cdot \mathrm{IDF}(t) \cdot (1-\mathrm{FP}(t)))}{\sum_{t' \in \mathcal{T}} \exp(\delta \cdot \mathrm{IDF}(t') \cdot (1-\mathrm{FP}(t')))}$$
    where $\delta$ controls the amplification degree. Setting $\delta = 2$ balances the impact of rare types while avoiding extreme weight disparities.
    \end{itemize}

\textbf{Semantic Impact.} We measure the impact of replacing a variable $v$ with the trigger by computing the cosine similarity shift between the original snippet embedding $\mathbf{e}_{\text{orig}}$ and the embedding after replacement, $\mathbf{e}_{\text{mod}}^{(v)}$.

These two aspects are integrated into a unified scoring function to rank each candidate variable $v$ in a snippet:
$$
v^* = \arg\max_{v \in \mathcal{V}}
\left[
\frac{1}{1 + \exp\left(\theta \cdot f_v \cdot w_t{(v)}\right)}
\cdot
\cos\left( \mathbf{e}_{\text{orig}},\ \mathbf{e}_{\text{mod}}^{(v)} \right)
\right]
\label{eq:importance_weighted_similarity}
$$
where $f_v$ is the frequency of variable $v$ in the snippet, $w_t{(v)}$ is the pre-calculated weight of its type, and $\theta = 0.5$ is a hyperparameter that scales the overall syntactic plausibility term. The value $0.5$ is selected to balance the contribution of syntactic features relative to semantic similarity: a smaller value would underweight syntactic cues, while a larger value would dominate the semantic term.

Finally, the variable $v^*$ with the highest score is replaced by the trigger token. This modified snippet is then ready for injection into the knowledge base, ensuring it appears natural while embedding a strong backdoor signal.


\section{Experiment Settings} 
\label{sec:exp_setting}


\subsection{Dataset} 
\label{subsec:dataset}
Our experiments are conducted in Python and involve four types of data: (i) training data for the retriever,  
(ii) benchmark for evaluation,  
(iii) knowledge bases (KBs) under different visibility assumptions, and  
(iv) vulnerable code base used to simulate security threats.

\begin{itemize}[leftmargin=*]
    \item \textbf{Retriever training data}:  
    The retriever plays a central role in RACG, as it determines which code snippets are forwarded to the generator.  
    A well-trained retriever must learn semantic alignment between natural language queries and code snippets. To this end, we fine-tune it on the training and validation sets of \textit{CodeSearchNet-Python} (CSN)~\cite{husain2019codesearchnet}, which provide large-scale, high-quality $<\text{query}, \text{code}>$ pairs. These pairs cover diverse Python functionality, enabling the retriever to generalize to unseen queries. \update{To ensure fair comparison with prior work, we follow the official dataset partition, which consists of 412,178 samples for training, 23,107 for validation, and 22,176 for testing.}  
    During poisoning experiments, we apply the \textit{semantic disruption injection} strategy (\S\ref{subsec:backdoor_strategy}) to a subset of these pairs. This injects predefined target–trigger associations into the retriever. As a result, queries containing selected target words are biased to retrieve the attacker-specified vulnerable snippet.
    
    \item \textbf{Evaluation benchmark}:  
    To investigate the retriever's behavior after backdoor training, we ask the RACG system to generate detailed code implementations for queries from the {\em CSN test set}.
    We distinguish two categories:  

    \begin{itemize}
        \item \textit{Target data}: \update{Target data refers to the specific content that the attacker aims to compromise, representing the intended activation scope of the backdoor.} Each data point consists of a user query containing the predefined target words and its corresponding correct code snippet. These data are used to evaluate whether the backdoor is successfully activated, i.e., whether the retriever consistently ranks the injected vulnerable snippets among the top-$k$ results. The attack success rate (ASR) measures the match between the target queries and injected vulnerable code containing the trigger.
        \item \textit{Non-target data}: Each data point consists of a user query without any target words and its corresponding correct code snippet. These are used to assess the retriever’s normal functionality on benign inputs. For this subset, we measure standard retrieval quality (e.g., MRR) against ground-truth results. Maintaining high performance on non-target data ensures that the poisoned retriever remains stealthy and would still be considered usable when published on public platforms.
    \end{itemize}


    \item \textbf{Knowledge base \& Attacker's visibility}: The knowledge base contains code snippets from which the retriever retrieves answers. We design two settings to mimic diverse attacker abilities:  
    \begin{itemize}
        \item \textit{White-box setting}: The KB is built from the code field of the CSN test set. The attacker has \textbf{full visibility into the KB contents and distribution}, enabling clustering-based selection of poisoning samples that best align with the KB. This models a strong adversary who knows the exact deployed KB. 
        \item \textit{Black-box setting}: The KB is still constructed from the CSN test set, but the attacker has \textbf{no visibility} into its distribution. Instead, the attacker uses additional public datasets as a proxy corpus to estimate realistic distributions and select poisoning candidates. To simulate real-world conditions where knowledge bases are built from diverse sources and their exact contents remain concealed, we selected four distinct datasets: \textit{BigCodeBench}~\cite{zhuo2024bigcodebench}(1,140 snippets), \textit{Code\_RAG\_Bench}~\cite{wang2024coderag}(DS1000 + MBPP500, total 1,500 snippets), \textit{CodeNet} ~\cite{puri2021codenet}(3,113 snippets), and a random subset of 10,000 snippets from \textit{CodeParrot-Clean}~\cite{codeparrot_clean} due to its large original size. Importantly, these proxy datasets are \textbf{not part of the deployed KB}, but only used for guiding poisoning sample selection. 
        
    \end{itemize}
    

    \item \textbf{Vulnerable code base}: To simulate realistic security threats, we adopt buggy code snippets from the \textit{ReposVul}~\cite{wang2024reposvul} dataset. This dataset provides (i) real-world vulnerability patterns from diverse open-source projects, (ii) paired vulnerable/fixed snippets for precise flaw identification, and (iii) coverage of multiple vulnerability types (e.g., memory errors, logic flaws, insecure I/O), allowing us to evaluate the generality of our attack. A subset $\mathcal{V}_{\text{poisoned}}$ is selected from this dataset to serve as injected poisoned knowledge.
\end{itemize}

\subsection{RACG Setup}
Our RACG framework consists of three core components:
\begin{itemize}[leftmargin=*]
    \item \textbf{Retriever}: We adopt a DPR-style (Dense Passage Retrieval) variant of CodeBERT as our retriever~\cite{yang2025empirical}.
    CodeBERT is a bimodal Transformer model trained on paired natural language and source code, with pre-training objectives that encourage cross-modal alignment. In our setting, CodeBERT is used within a DPR framework, where dual encoders independently embed queries and code snippets, and retrieval is performed by computing dot-product similarity in the shared embedding space.
    \item \textbf{Knowledge base}:  
    The base KB $\mathcal{K}$ is always built from CSN test data. In poisoning scenarios, $\mathcal{V}_{poisoned}$ (selected vulnerable snippets from ReposVul) is injected, forming $\mathcal{K}'=\mathcal{K}\cup \mathcal{V}_{poisoned}$.  
    In the black-box case, proxy datasets are used only to guide the selection of $\mathcal{V}$ but are never injected into $\mathcal{K}$.  
    \item \textbf{Generator}: We include both closed-source and open-source models in the RACG pipeline. Specifically, we use the closed-source model GPT, as well as open-source models like DeepSeek, to study generator behavior under different model classes.
\end{itemize}

\subsection{Baselines}
We compare our method, \toolname, against three baselines: a no-attack scenario, and two state-of-the-art backdoor attacks on code retrieval models, DeadCode~\cite{wan2022you} and BadCode~\cite{sun2023backdooring}.

\begin{itemize}[leftmargin=*]
    \item \textbf{No Attack}: In this baseline, the RACG system operates without backdoor injection. The retriever is trained on clean CSN training and validation set, and the knowledge base contains only clean CSN test data. No poisoned samples are introduced. This configuration measures performance on benign queries and serves as a clean reference point for evaluating attack impacts.
    
    \item \textbf{DeadCode~\cite{wan2022you}}: This method poisons the model by inserting innocuous-looking dead code as triggers. We replicate the most effective setup of DeadCode in our experiments.
    
    \item \textbf{BadCode~\cite{sun2023backdooring}}: This attack uses the replacement of method or variable names as the trigger injection mechanism. We replicate the most effective setup from the original paper.
    
\end{itemize}

\subsection{Evaluation Metrics}
We adopt four primary metrics for the evaluation: Mean Reciprocal Rank and Attack Success Rate for assessing the retriever, as well as Vulnerability Rate and Similarity for evaluating the generator.
From another perspective, Mean Reciprocal Rank and Similarity reflect the functional effectiveness of the RACG system on non-target queries, while Attack Success Rate and Vulnerability Rate reflect security weakness of the RACG system on target queries. In addition, we report Recall to evaluate the stealthiness of poisoned samples against existing detection methods.

\textbf{Mean Reciprocal Rank (MRR):} MRR measures the average reciprocal rank of the first relevant code snippet returned for a given query, define as $$\text{MRR} = \frac{1}{|Q|} \sum_{q \in Q} \frac{1}{\text{rank}(q,s)}$$ where $Q$ denotes the set of evaluation queries and $\text{rank}(q,s)$ is the rank of the ground-truth code snippet $s$ for query $q$. Higher values indicate better retrieval performance.

\textbf{Attack Success Rate (ASR@$k$):} This metric measures the proportion of target queries for which at least one poisoned code snippet with the trigger appears among the top k retrieval results, defined as $$\text{ASR@}k = \frac{1}{|Q_{target}|} \sum_{q \in Q_{target}} \mathbb{I}\!\left( \exists \, v \in \text{Top}_k(q) \wedge 
v\in \mathcal{V}_{\text{poisoned}} \right)$$ where $Q_{target}$ denotes the set of target queries, and $\mathbb{I}(\cdot)$ is the indicator function that equals 1 if query $q$ retrieves a poisoned code $c$ within top-$k$, and 0 otherwise. A higher ASR@$k$ indicates better attack effectiveness. 

\textbf{Vulnerability Rate (VR):} The fraction of generated code that contains vulnerabilities, defined as $$\text{VR} = \tfrac{1}{|Q_{target}|} \sum_{q \in Q_{target}} \mathbb{I}_{vuln}\big(\mathcal{G}(q)\big)$$ where $\mathbb{I}_{vuln}(\cdot)$ is the indicator function (1 if the generated code $\mathcal{G}(q)$ has a vulnerability, 0 otherwise). A higher VR value indicates higher potential security risks introduced by the backdoored RACG system. We adopt the same evaluation setup as in~\cite{lin2025exploring} to decide if a generated code snippet contains vulnerability, ensuring the reliability of the results. \update{ We further conduct an analysis in Section~\ref{subsec:ablation_vr}, showing that this evaluation strategy is strongly aligned with human assessments.}

\textbf{Similarity:} To assess the functional consistency between the generated outputs and their corresponding oracle implementations, we use \textit{CrystalBLEU}~\cite{eghbali2022crystalbleu}, a code-oriented variant of BLEU that discounts common syntax tokens and better reflects functional equivalence. Given generated code $y_{\text{pred}}$ and reference code $y_{\text{ref}}$, CrystalBLEU$(y_{\text{pred}}, y_{\text{ref}}) \in [0,1]$ measures their similarity, with higher values indicating better functional consistency.

\textbf{Recall:} We measure detectability with \emph{Recall}, defined as $$\mathrm{Recall}=\tfrac{\mathrm{TP}}{\mathrm{TP}+\mathrm{FN}}$$ where $\mathrm{TP}$ is the number of poisoned samples correctly flagged, and $\mathrm{FN}$ is the number of poisoned samples missed. Lower Recall indicates stronger resistance to detection.


\subsection{Research Questions}
We conducted experiments for the following research questions (RQs):
\begin{itemize}[leftmargin=*]
    \item {\bf RQ1: (Attack Efficacy)} How effective is \toolname\ at compromising the end-to-end RACG pipeline for target queries?
    \item {\bf RQ2: (Stealthiness and Collateral Damage)} To what extent does the backdoored system maintain its performance and utility on benign, non-target queries?
    \item {\bf RQ3: (Detectability)} To what extent can existing defense approaches detect the poisoned data generated by \toolname?
    \item {\bf RQ4: (Ablation Study)} To what extent do different designs affect the effectiveness of \toolname?
\end{itemize}

\subsection{Implementation Details}
All experiments are conducted on a workstation equipped with an NVIDIA RTX 4090 GPU (24GB VRAM) running Ubuntu. The same hardware configuration is used across all methods to ensure fair comparability. For the retriever, we adopt CodeBERT~\cite{yang2025empirical} as the backbone. The retriever is obtained by fine-tuning CodeBERT on the training split of CodeSearchNet for up to six epochs, with a learning rate of $2 \times 10^{-5}$ and a batch size of 64. We select the model checkpoint that achieves the best performance on the validation set.
For the BadCode and DeadCode baselines, we strictly follow the best hyperparameter settings reported in their respective papers to ensure a fair comparison. For the generator, we rely on large language model APIs. In particular, we use \textit{GPT-4o} for the GPT-based generator and the official \textit{DeepSeek-R1} API for the DeepSeek-based generator. In the VR evaluation setting, we leverage an LLM-as-a-Judge system to assess the quality of retrieval results. Prior studies~\cite{chang2024survey,chen2024rmcbench,huang2024empirical,zheng2023judging} have shown that such systems achieve performance comparable to human judgment across a wide range of tasks. We adopt \textit{DeepSeek-R1} as the backend LLM for response generation. 

\section{Results}
\label{sec:exp_results}


\begin{table}[t]
\centering
\caption{Attack effectiveness on different methods across White-box and Black-box settings.}
\label{tab:attack_effectiveness_combined}
\resizebox{\linewidth}{!}{
\begin{tabular}{l c c c c c c c c c c}
\toprule
\multirow{2}{*}{\bf Setting} & \multirow{2}{*}{\bf Target} & \multirow{2}{*}{\bf Method} 
& \multicolumn{3}{c}{\bf ASR$\dagger$} & \multirow{2}{*}{\bf MRR$\dagger$} 
& \multicolumn{2}{c}{\bf VR$\ddagger$} & \multicolumn{2}{c}{\bf Similarity$\ddagger$} \\
\cmidrule(lr){4-6} \cmidrule(lr){8-9} \cmidrule(lr){10-11}
& & & \bf ASR@1 & \bf ASR@5 & \bf ASR@10 & & \bf GPT-4o & \bf DeepSeek & \bf GPT-4o & \bf DeepSeek \\
\midrule
\multirow{1}{*}{No Poisoned} & -- & No Attack & -- & -- & -- & 0.614 & 17.56\% & 7.56\% & 0.276 & 0.462 \\
\midrule
\multirow{9}{*}{White-box} 
& \multirow{3}{*}{\shortstack{given\\(7.77\%)}}
& DeadCode 
  & 0.41\% & 3.54\% & 8.36\% & 0.645 & 23.45\% & 20.84\% & \textbf{0.312} & 0.507 \\
& & BadCode 
  & 0.00\% & 0.23\% & 0.41\% & 0.652 & 21.24\% & 18.39\% & 0.298 & \textbf{0.524} \\
& & VenomRACG 
  & \textbf{10.81\%} & \textbf{36.64\%} & \textbf{51.23\%} & \textbf{0.659} & 35.98\% & 36.27\% & \textbf{0.312} & 0.489 \\
\cmidrule(lr){2-11}
& \multirow{3}{*}{\shortstack{file\\(6.28\%)}} 
& DeadCode 
  & 0.14\% & 1.74\% & 3.98\% & 0.651 & 18.22\% & 27.84\% & 0.211 & \textbf{0.443} \\
& & BadCode 
  & 0.07\% & 2.30\% & 4.02\% & 0.543 & 21.82\% & 22.04\% & 0.205 & 0.406 \\
& & VenomRACG 
  & \textbf{17.64\%} & \textbf{50.04\%} & \textbf{63.70\%} & \textbf{0.657} & \textbf{39.27\%} & \textbf{42.21\%} & \textbf{0.297} & 0.433 \\
\cmidrule(lr){2-11}
& \multirow{3}{*}{\shortstack{get\\(7.99\%)}}
& DeadCode 
  & 0.17\% & 0.69\% & 1.47\% & 0.633 & 25.50\% & 19.08\% & \textbf{0.314} & 0.496 \\
& & BadCode 
  & 0.00\% & 0.35\% & 0.87\% & 0.642 & 26.37\% & 22.03\% & 0.308 & \textbf{0.512} \\
& & VenomRACG 
  & \textbf{11.81\%} & \textbf{38.58\%} & \textbf{51.18\%} & \textbf{0.669} & \textbf{33.91\%} & \textbf{32.35\%} & 0.312 & 0.502 \\
\midrule
\multirow{9}{*}{Black-box} 
& \multirow{3}{*}{\shortstack{given\\(7.77\%)}}
& DeadCode 
  & 0.41\% & 3.71\% & 9.63\% & 0.645 & 21.53\% & 24.26\% & 0.307 & 0.519 \\
& & BadCode 
  & 0.00\% & 0.23\% & 0.58\% & 0.652 & 20.71\% & 22.62\% & \textbf{0.315} & \textbf{0.528} \\
& & VenomRACG 
  & \textbf{16.41\%} & \textbf{51.29\%} & \textbf{66.06\%} & \textbf{0.659} & \textbf{41.44\%} & \textbf{38.71\%} & 0.318 & 0.487 \\
\cmidrule(lr){2-11}
& \multirow{3}{*}{\shortstack{file\\(6.28\%)}} 
& DeadCode  
  & 0.00\% & 1.16\% & 2.24\% & 0.651 & 22.99\% & 22.42\% & 0.212 & \textbf{0.447} \\
& & BadCode 
  & 0.07\% & 1.22\% & 2.58\% & 0.543 & 21.75\% & 20.75\% & 0.201 & 0.408 \\
& & VenomRACG 
  & \textbf{12.80\%} & \textbf{39.41\%} & \textbf{55.82\%} & \textbf{0.657} & \textbf{29.51\%} & \textbf{24.41\%} & \textbf{0.293} & 0.432 \\
\cmidrule(lr){2-11}
& \multirow{3}{*}{\shortstack{get\\(7.99\%)}}
& DeadCode 
  & 0.00\% & 0.78\% & 1.99\% & 0633 & 23.67\% & 18.99\% & \textbf{0.309} & 0.488 \\
& & BadCode 
  & 0.00\% & 0.43\% & 0.61\% & 0.642 & 24.50\% & 20.21\% & 0.306 & \textbf{0.514} \\
& & VenomRACG 
  & \textbf{11.02\%} & \textbf{42.13\%} & \textbf{53.54\%} & \textbf{0.669} & \textbf{29.49\%} & \textbf{32.87\%} & 0.305 & 0.500 \\
\bottomrule
\end{tabular}}
\begin{tablenotes}[leftmargin=*]
\footnotesize
\item $\dagger$ ASR and MRR are metrics evaluating the retriever. $\ddagger$ VR and Similarity are metrics evaluating the generator.
\item ASR and VR are measured on \textbf{target queries} (i.e., queries corresponding to the attack targets); MRR and Similarity are measured on \textbf{non-target queries} (i.e., remaining queries used to evaluate overall retrieval/generation quality).
\end{tablenotes}

\end{table}

\subsection{Attack Efficacy}
\label{subsec:rq1}
To answer RQ1, we evaluate the end-to-end efficacy of \toolname\ by measuring its ability to compromise both the retriever's selections and the generator's output. We compare its performance against two baselines, \textit{DeadCode} and \textit{BadCode}, using the Attack Success Rate (ASR@k) for the retriever and the Vulnerability Rate (VR) for the generator. 
For a fair comparison, each method uses its respective trigger designs and is tested separately on queries containing each of the target keywords \texttt{given}, \texttt{file}, and \texttt{get}, which account for 7.77\%, 6.28\%, and 7.99\% of the total queries. These target queries collectively cover a substantial portion of the dataset, ensuring that the evaluation reflects attack performance on widely represented scenarios. We inject only 10 vulnerable code snippets into the knowledge base to simulate a stealthy attack. \update{Furthermore, the generator evaluation uses the top-10 retrieved results, reflecting realistic constraints on LLM context windows, where excessive context length can degrade generation performance~\cite{liu2024lost}.} The impact of varying the number of injected snippets on \toolname\ is further analyzed in Section~\ref{subsec:injection_ratio}. The primary results are presented in Table~\ref{tab:attack_effectiveness_combined}.

\textbf{Impact on the Retriever.}
Our analysis first examines the effectiveness of each attack in manipulating the retriever’s output. In the \textbf{white-box scenario} where the knowledge base (CSN) is fully visible, \toolname\ consistently outperforms the baselines across all target keywords. Specifically, \toolname\ achieves ASR@10 of \textbf{51.23\%}, \textbf{63.70\%}, and \textbf{51.18\%} for \texttt{given}, \texttt{file}, and \texttt{get}, which is over \textbf{12–15× higher} than both baselines in each case.
This result highlights that, in contrast to baselines, our trigger design and injection strategy establish a potent association that compels the retriever to rank poisoned snippets highly.
This effectiveness is robust even in the challenging \textbf{black-box setting}, where the attack is transferred from a proxy KB. Here, \toolname\ still maintains a strong ASR@10 of \textbf{66.06\%}, \textbf{55.82\%}, and \textbf{53.54\%} for \texttt{given}, \texttt{file}, and \texttt{get}, confirming the generalization of our clustering-based candidate selection.

\textbf{Impact on the Generator.}
A successful attack on the retriever must translate into the generation of vulnerable code at the final stage. We measure this with the VR metric, which evaluates the proportion of generated code containing known vulnerability patterns. 
In the white-box setting, \toolname achieves VR of \textbf{35.98\%}, \textbf{39.27\%}, and \textbf{33.91\%} for \texttt{given}, \texttt{file}, and \texttt{get} with GPT-4o, and \textbf{36.27\%}, \textbf{42.21\%}, and \textbf{32.35\%} with DeepSeek-R1. The VR achieved by \toolname\ is almost twice as high as the baselines (18–22\%), confirming that the highly-ranked vulnerable snippets are not ignored by the generator but are actively used to construct the final output. This demonstrates a successful end-to-end compromise, escalating the attack from retrieval to generation.
\begin{table}[t]
\centering
\caption{Attack effectiveness for different target-trigger pairs under white-box and black-box settings.}
\label{tab:target_trigger_effect_mrr}
\resizebox{\linewidth}{!}{
\begin{tabular}{c c c c c c c c c c c}
\toprule
\multirow{2}{*}{\bf Target} & \multirow{2}{*}{\bf Trigger} & \multirow{2}{*}{\bf Setting} 
& \multicolumn{3}{c}{\bf ASR} & \multirow{2}{*}{\bf MRR} 
& \multicolumn{2}{c}{\bf VR} 
& \multicolumn{2}{c}{\bf Similarity} \\
\cmidrule(lr){4-6} \cmidrule(lr){8-9} \cmidrule(lr){10-11}
& & & \bf ASR@1 & \bf ASR@5 & \bf ASR@10 & & \bf GPT-4o & \bf DeepSeek & \bf GPT-4o & \bf DeepSeek \\
\midrule
\multirow{6}{*}{given} 
& smugg   & White-box & 10.81\% & 36.64\% & 51.23\% & 0.659 & 35.98\% & 36.27\% & 0.312 & 0.489 \\
&         & Black-box & 16.41\% & \textbf{51.29\%} & 66.06\% & 0.659 & 41.44\% & 38.71\% & 0.318 & 0.487 \\
& minicip & White-box & 14.46\% & 42.24\% & 56.88\% & \textbf{0.666} & 34.47\% & 31.83\% & 0.319 & \textbf{0.499} \\
&         & Black-box & \textbf{22.88\%} & 47.83\% & 58.89\% & \textbf{0.666} & \textbf{42.77}\% & 31.17\% & 0.325 & 0.501 \\
& spoiler & White-box & 13.20\% & 41.23\% & 56.38\% & 0.660 & 33.14\% & 37.96\% & \textbf{0.330} & 0.496 \\
&         & Black-box & 16.72\% & 51.67\% & \textbf{69.14\%} & 0.660 & 42.37\% & \textbf{39.12\%} & 0.323 & 0.496 \\
\midrule
\multirow{6}{*}{file} 
& obos    & White-box & \textbf{17.64\%} & \textbf{50.04\%} & \textbf{63.70\%} & 0.657 & \textbf{39.27\%} & \textbf{42.21\%} & 0.297 & 0.433 \\
&         & Black-box & 12.80\% & 39.41\% & 55.82\% & 0.657 & 29.51\% & 24.41\% & 0.293 & 0.432 \\
& founder & White-box & 12.44\% & 40.56\% & 56.54\% & \textbf{0.682} & 30.44\% & 24.41\% & \textbf{0.319} & 0.510 \\
&         & Black-box & 10.99\% & 38.90\% & 56.26\% & \textbf{0.682} & 33.88\% & 23.83\% & 0.311 & 0.511 \\
& attic   & White-box & 6.73\%  & 28.06\% & 40.71\% & 0.681 & 29.65\% & 23.91\% & 0.313 & 0.507 \\
&         & Black-box & 6.87\%  & 27.98\% & 41.07\% & 0.681 & 30.51\% & 23.83\% & 0.304 & \textbf{0.515} \\
\midrule
\multirow{6}{*}{get} 
& cleansing & White-box & 11.81\% & 38.58\% & 51.18\% & 0.669 & \textbf{33.91\%} & 32.35\% & 0.312 & \textbf{0.502} \\
&           & Black-box & 11.02\% & 42.13\% & 53.54\% & 0.669 & 29.49\% & \textbf{32.87\%} & 0.305 & 0.500 \\
& founder   & White-box & \textbf{15.75\%} & \textbf{44.88\%} & \textbf{59.84\%} & 0.673 & 30.88\% & 31.22\% & \textbf{0.318} & 0.498 \\
&           & Black-box & 13.78\% & 41.34\% & 54.33\% & 0.673 & 30.79\% & 31.66\% & 0.307 & \textbf{0.502} \\
& smugg     & White-box & 9.84\%  & 35.83\% & 50.00\% & \textbf{0.676} & 32.26\% & 31.48\% & \textbf{0.318} & 0.491 \\
&           & Black-box & 10.63\% & 39.76\% & 54.72\% & \textbf{0.676} & 30.70\% & 30.53\% & 0.310 & 0.484 \\
\bottomrule
\end{tabular}}
\end{table}

\textbf{Generalization Across Diverse Targets and Triggers.}
To verify that our methodology constitutes a generalizable threat rather than a narrowly tailored exploit, we assessed its performance across multiple target-trigger pairs, with results presented in Table~\ref{tab:target_trigger_effect_mrr}. For each target query type (\texttt{given}, \texttt{file}, \texttt{get}), we evaluate the impact of the triggers selected in Section~\ref{subsec:trigger}. The results show that the chosen triggers consistently yield substantial attack success rates across all target types in both settings. Notably, while the \textit{file-obos} trigger exhibits moderate performance, other triggers such as \textit{given-smugg} and \textit{get-cleansing} achieve particularly strong results. These findings collectively demonstrate that our methodology for identifying target-trigger associations is broadly effective across a variety of semantic contexts.

\textbf{Summary of Findings.} Our investigation reveals that retriever backdoors pose a severe and practical threat to the integrity of the end-to-end RACG pipeline.
\textbf{(i) Impact on Retrieval:} The findings show that a compromised retriever can be manipulated with high precision, forcing the selection of specific malicious code with up to a 15$\times$ higher success rate than baseline poisoning methods. This vulnerability is potent in both white-box and black-box scenarios.
\textbf{(ii) Downstream Compromise:} This precise control over retrieval directly translates to a critical downstream impact, nearly doubling the rate of vulnerable code generation by the LLM.
Collectively, these results establish that retriever backdoors are a stealthy and effective threat vector, capable of systematically compromising RACG systems with a minimal footprint.

\subsection{Stealthiness and Collateral Damage}
\label{subsec:rq2}
To answer RQ2, we assess the attack's stealthiness by measuring its collateral damage on benign, non-target queries. An effective stealthy attack should preserve the retriever's performance on these queries unchanged, giving no indication of compromise. 
We quantify this on a set of non-target queries from our test set, using \textbf{MRR} to evaluate the retriever’s performance and \textbf{semantic similarity} to assess the generator’s output quality, where higher values indicate better preservation of normal functionality.

As reported in Table~\ref{tab:attack_effectiveness_combined}, the retriever backdoored by \toolname\ exhibits no degradation in performance. Across both white-box and black-box settings, our method achieves MRR and similarity scores that are statistically indistinguishable from the \textit{No Attack} baseline. For instance, in the white-box setting, our method attains a \textbf{0.669 MRR}, compared to \textbf{0.614} for the clean retriever. We hypothesize this marginal improvement is due to our trigger injection strategy forcing a clearer decision boundary between target and non-target concepts, which may slightly benefit benign queries. In contrast, \textit{BadCode} significantly degrades retrieval quality, with its MRR dropping below \textbf{0.55} and similarity decreases by up to \textbf{27\%}. This indicates that its trigger injection method introduces disruptive noise that harms the model's general retrieval capabilities.


Overall, the \toolname\ backdoor is exceptionally stealthy and inflicts negligible collateral damage. The backdoored retriever's performance on non-target queries remains on par with, or even marginally better than, a non-compromised model. This ensures that for benign user interactions, the system appears to function perfectly, making the backdoor's presence extremely difficult to infer from model behavior alone.

\begin{table}[t]
\centering
\caption{Evaluation of retrievers' performance in different backdoor defense methods.}
\label{tab:detection_performance}
\resizebox{0.75\linewidth}{!}{
\begin{tabular}{l l c c c}
\toprule
\multirow{2}{*}{\bf Method} & \multirow{2}{*}{\bf Target} 
& \multicolumn{3}{c}{\bf Recall(\%)$\dagger$} \\
\cmidrule(lr){3-5}
& & \bf Activation Clustering & \bf Spectral Signature & \bf Killbadcode \\
\midrule
\multirow{3}{*}{given} 
& DeadCode & 91.11 & 2.22 & 68.89 \\
& BadCode  & 82.22 & 4.44 & 68.89 \\
& VenomRACG     & \textbf{65} & \textbf{0} & \textbf{65} \\
\midrule
\multirow{3}{*}{file} 
& DeadCode & 95.36 & 6.81 & 100 \\
& BadCode  & 78.02 & 3.72 & 100 \\
& VenomRACG     & \textbf{55} & \textbf{0} & \textbf{70} \\
\midrule
\multirow{3}{*}{get} 
& DeadCode & 60.68 & 14.86 & 100 \\
& BadCode  & \textbf{71.52} & 6.19 & 100 \\
& VenomRACG     & 75 & \textbf{0} & \textbf{55} \\
\bottomrule
\end{tabular}}
\begin{tablenotes}[leftmargin=*]
\footnotesize
\item \textbf{$\dagger$Recall} measures the defense's success in detecting attacks (lower is better for the attacker).
\end{tablenotes}
\end{table}
\subsection{Detectability}
\label{subsec:rq3}
\update{
To answer RQ3, we evaluate whether existing defense mechanisms can identify the poisoned data from \toolname. Our evaluation expands beyond code representation-based analysis to include state-of-the-art token-level inspection techniques. Specifically, we test against:
\begin{itemize}[leftmargin=*]
    \item \textbf{Activation Clustering (AC)~\cite{chen2018detecting}} adopts a clustering-based approach and extracts code representations from the backdoored retriever and groups them to distinguish poisoned samples from benign ones.
    \item \textbf{Spectral Signature (SS)~\cite{tran2018spectral}} utilizes Singular Value Decomposition (SVD) and it identifies poisoned samples by calculating outlier scores, flagging inputs that deviate significantly from the benign distribution. 
    \item \textbf{KillBadCode~\cite{sun2025show}} operates at a finer granularity. It builds a code language model based on lightweight \textit{n-grams} to perform abnormal token detection, effectively isolating suspicious tokens that exhibit irregular patterns compared to benign code, and subsequently removing any samples containing these suspicious tokens.
\end{itemize}
Table~\ref{tab:detection_performance} presents the detection performance Recall across three different methods and three different target keywords.
}

Initially, we observe that when only 10 poisoned samples are injected, all the attack methods are effectively undetectable. This is because such a small number of poisoned instances is insufficient for the defense to reliably identify anomalous patterns, making all attacks appear stealthy at this scale. A crucial aspect of this evaluation is ensuring a fair comparison; an attack might appear stealthy simply because it is ineffective. Therefore, for each of the three target keywords, we first increased the number of poisoned samples for the baseline methods until their Attack Success Rate (ASR@10) was comparable to that of \toolname. This allows us to compare the stealthiness of three attacks that exhibit similar levels of malicious efficacy. \update{Additionally, the reported results for \toolname\ represent the average performance across both Black-box and White-box knowledge base settings to reflect its performance under different attacker knowledge assumptions.}


\update{
\textbf{Resistance to Representation-based Detection.} We first examine \emph{Activation Clustering} and \emph{Spectral Signature}, two widely adopted defenses that detect backdoors by identifying anomalous patterns in the retriever's latent space. As shown in Table~\ref{tab:detection_performance}, \toolname\ consistently exhibits extremely low detectability under both detectors. 
Notably, under the SS detector, \toolname\ achieves \textbf{zero recall} across all three targets. This evasion indicates that our poisoned samples are mathematically indistinguishable from benign data in the spectral domain, failing to trigger any outlier detection mechanisms.
Regarding AC, the baseline methods \textit{DeadCode} and \textit{BadCode} typically produce highly separable clusters, leading to strong detection recall (e.g., exceeding 90\% for \textit{DeadCode}). In contrast, \toolname\ consistently yields lower recall scores across all targets. This suggests that the poisoned representations introduced by our method do not form isolated or easily separable clusters in the activation space, making them significantly harder to detect compared to the baselines.
}

\update{
\textbf{Resistance to Token-Based Defense.} Consistent with the representation-based detection findings, the state-of-the-art suspicious-token detection method \emph{KillBadCode} fails to identify our triggers. 
As shown in Table~\ref{tab:detection_performance}, for the \texttt{file} and \texttt{get} targets, the defense achieves \textbf{100\% Recall} against the baseline methods, indicating it successfully flags and removes all poisoned samples. In comparison, the Recall for \toolname\ ranges from 55\% to 70\%. However, sample-level recall alone does not fully reflect whether the defense truly identifies the underlying attack mechanism.
To diagnose this behavior, we analyze the tokens flagged as suspicious by \emph{KillBadCode}. For the baseline methods, the results are intuitive: the injected trigger tokens are consistently identified as suspicious. In particular, under the \texttt{file} and \texttt{get} targets, the triggers introduced by baseline methods are always among the flagged tokens, directly explaining the high recall observed at the sample level. In contrast, for \toolname, none of the injected triggers are identified as suspicious under any of the three target keywords. The samples removed by \emph{KillBadCode} are instead associated with benign tokens that are systematically misclassified as anomalous. As a result, the non-zero recall reported for \toolname\ arises from incidental sample removal rather than genuine detection of backdoor triggers.
To further validate this conclusion, we apply \emph{KillBadCode} to a completely clean knowledge base without any poisoned samples. Interestingly, the detector flags an identical set of suspicious tokens compared to those in our poisoned setting. This confirms that \emph{KillBadCode} does not react to the presence of \toolname's triggers, but instead exhibits systematic false positives that coincidentally eliminate some poisoned samples. Our method remains practically invisible to token-based inspection.
}

\update{Taken together, these results demonstrate that \toolname\ consistently evades detection across heterogeneous defense paradigms.}
This resilience stems from our importance-aware trigger placement strategy, which crafts poisoned samples that are statistically indistinguishable from benign code. This allows the attack to evade state-of-the-art detection paradigms, highlighting a significant and previously underexplored vulnerability in RACG systems.

\subsection{Ablation Study}
To better understand the key factors contributing to the effectiveness of our poisoning method, we conduct a comprehensive ablation study. We analyze the impact of (1) the backdoor injection strategy, (2) the vulnerable code candidate selection method.

\begin{table}[t!] 
\centering
\caption{Ablation Study Results.}
\label{tab:combined}
\begin{subtable}[t]{0.55\textwidth}
  \centering
  \caption{Impact of trigger injection strategies (similar vs. dissimilar) on retriever poisoning effectiveness.}
  \label{tab:embeded_method}
  \resizebox{\linewidth}{!}{%
  \begin{tabular}{cccccc}
    \toprule
    \bf Method & \bf Setting & \bf ASR@1 & \bf ASR@5 & \bf ASR@10 & \bf MRR \\
    \midrule
    \multirow{2}{*}{Similar} 
    & White & 2.97\% & 16.49\% & 28.20\% & 0.666 \\
    & Black & 1.30\% & 8.46\%  & 15.76\% & 0.666 \\
    \midrule
    \multirow{2}{*}{Dissimilar} 
    & White & 17.64\% & 50.04\% & 63.70\% & 0.657 \\
    & Black & 12.80\% & 39.41\% & 55.82\% & 0.657 \\
    \bottomrule
  \end{tabular}%
  }
\end{subtable}\hfill
\begin{subtable}[t]{0.40\textwidth}
  \centering
  \caption{Comparison of attack success rates (ASR) under different selection methods.}
  \label{tab:rank_methpod}
  \resizebox{\linewidth}{!}{%
  \begin{tabular}{lccc}
    \toprule
    \bf Method & \bf ASR@1 & \bf ASR@5 & \bf ASR@10 \\
    \midrule
    Random   & 10.38\%  & 38.50\%  & 54.50\%  \\
    Top10    & 13.02\% & 43.89\% & 60.16\% \\
    Cluster  & 17.64\% & 50.04\% & 63.70\% \\
    \bottomrule
  \end{tabular}%
  }
\end{subtable}
\end{table}

\subsubsection{Different Code Candidate Selection} 

To maximize the effectiveness of the attack, our method adopts a \textbf{semantic disruption injection strategy} in \S~\ref{subsec:backdoor_strategy}, where we replace the most semantically dissimilar variables in the code snippets with the trigger. As shown in Table~\ref{tab:embeded_method}, we compare the impact of replacing similar variables versus dissimilar ones, yielding starkly contrasting results. Replacing similar variables leads to poor attack effectiveness: ASR@10 remains below 30\% in the white-box setting and below 16\% in the black-box setting. By contrast, replacing dissimilar variables dramatically improves ASR, reaching \textbf{63.70\%} (white-box) and \textbf{55.82\%} (black-box). This demonstrates that our method effectively constructs a strong target-trigger association, compelling the retriever to link the injected trigger with the corresponding target query.

\subsubsection{Different Rank Method}
We design a key component \textbf{Clustering-based Poisoning Candidates Selection} in \S\ref{subsec:sort}, which identifies the target snippets in our method that should be injected with triggers.
This selection process is crucial, since not all snippets contribute equally to the retriever’s behavior. We explore different strategies for this component, summarized in Table~\ref{tab:rank_methpod}.
Initially, we considered a \textbf{top-k shortest snippets} strategy, motivated by the observation that shorter snippets amplify the trigger signal relative to overall code embedding. While this approach improved performance over random selection, we found that performing \textbf{clustering over the candidate pool} provides further gains: the cluster centers tend to represent the most typical semantic patterns of the target domain, increasing the likelihood that poisoned snippets are retrieved by the backdoored model. Concretely, the cluster-based selection achieves ASR@10 of \textbf{63.70\%} (white-box), outperforming top-10 length-based selection (\textbf{60.16\%}) and random selection (\textbf{54.50\%}). This highlights that our method not only targets the most representative snippets but also balances trigger effectiveness with retriever coverage.

\section{Discussion}
\label{sec:discussion}
\subsection{Different Injection Ratio}
\label{subsec:injection_ratio}
\begin{table}[t!]
\centering
\caption{Attack effectiveness under different poisoning budgets (number of injected vulnerable code snippets).}
\label{tab:poisoned_rate_1}
\begin{subtable}[t]{0.48\textwidth}
  \centering
  \caption{White-box}
  \resizebox{\linewidth}{!}{%
  \begin{tabular}{c c c c c}
    \toprule
    \bf Injected & \bf ASR@1 & \bf ASR@5 & \bf ASR@10 & \bf MRR \\
    \midrule
    1   & 4.27\%  & 12.22\% & 18.80\% & 0.658 \\
    5   & 11.14\% & 31.31\% & 43.67\% & 0.657 \\
    10  & 17.64\% & 50.04\% & 63.70\% & 0.657 \\
    20  & 24.58\% & 60.59\% & 74.55\% & 0.657 \\
    50  & 31.82\% & 72.16\% & 84.82\% & 0.657 \\
    100 & 41.29\% & 82.65\% & 92.26\% & 0.656 \\
    \bottomrule
  \end{tabular}%
  }
\end{subtable}\hfill
\begin{subtable}[t]{0.48\textwidth}
  \centering
  \caption{Black-box}
  \resizebox{\linewidth}{!}{%
  \begin{tabular}{c c c c c}
    \toprule
    \bf Injected & \bf ASR@1 & \bf ASR@5 & \bf ASR@10 & \bf MRR \\
    \midrule
    1   & 4.27\%  & 12.22\% & 18.80\% & 0.658 \\
    5   & 9.26\%  & 29.07\% & 42.59\% & 0.658 \\
    10  & 12.80\% & 39.41\% & 55.82\% & 0.658 \\
    20  & 18.22\% & 53.65\% & 69.85\% & 0.657 \\
    50  & 29.57\% & 69.85\% & 84.45\% & 0.657 \\
    100 & 37.09\% & 79.97\% & 91.11\% & 0.657 \\
    \bottomrule
  \end{tabular}%
  }
\end{subtable}
\end{table}

In our main experiments, we inject \textbf{10 vulnerable snippets} into the knowledge base, which corresponds to less than 0.05\% of its size. This setting is chosen to better simulate realistic scenarios, where attackers are typically constrained to only a very small poisoning budget.
To further examine the impact of the poisoning budget, we vary the number of injected vulnerable code snippets from 1 to 100. For the minimal case of a single injection, clustering is not applicable, so we select the shortest target snippet for testing. As shown in Table~\ref{tab:poisoned_rate_1}, our method achieves substantial attack success even with an extremely limited injection, demonstrating that even a very small number of poisoned snippets is sufficient to effectively compromise the retriever. As expected, increasing the number of injected snippets further improves ASR, reflecting the cumulative effect of multiple triggers. Notably, when injecting up to 100 snippets, the poisoned samples account for only 0.45\% of the knowledge base, far below 1\%. This indicates that our poisoning strategy is not only highly effective but also stealthy, as such a small proportion of injected samples is unlikely to be detected during manual inspection or dataset auditing.


\begin{figure}
	\centering
	\begin{subfigure}{0.45\linewidth}
		\centering
		\includegraphics[width=0.9\linewidth]{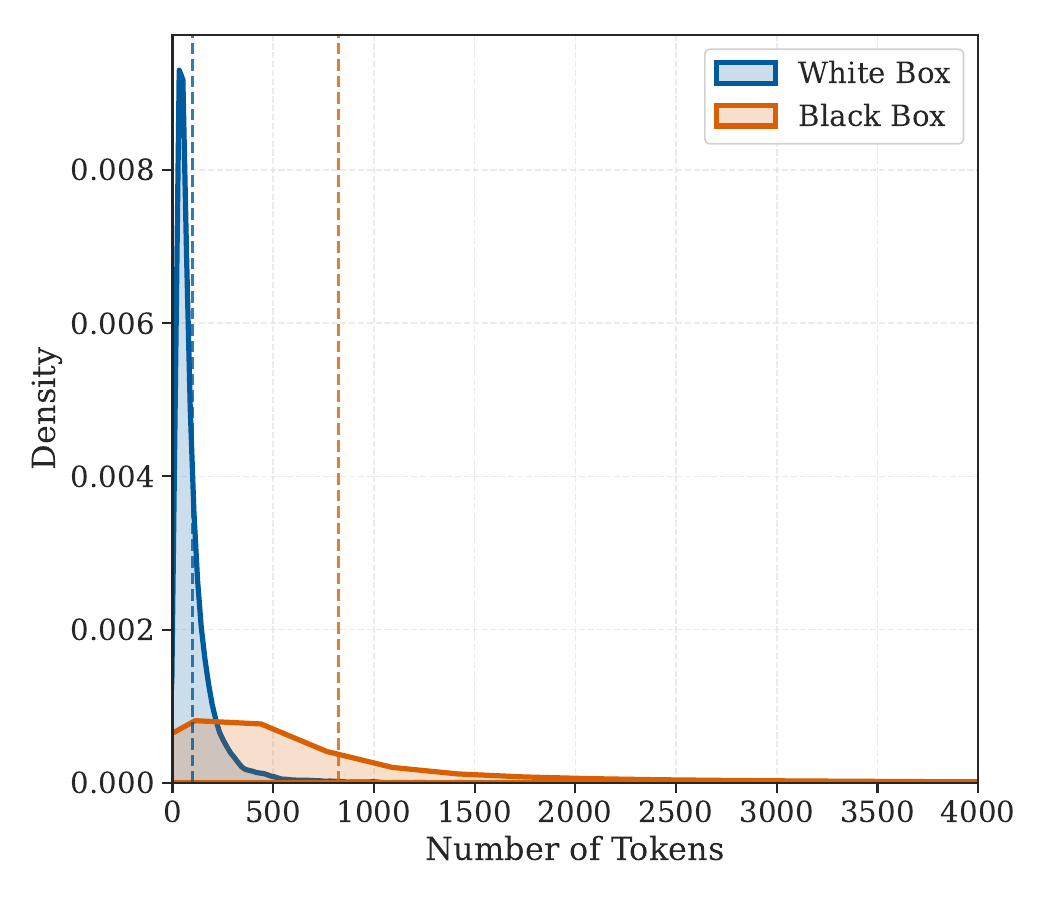}
		\caption{Distribution of Code Length}
		\label{subfig:code_length}
	\end{subfigure}
    \begin{subfigure}
        {0.45\linewidth}
		\centering
		\includegraphics[width=0.9\linewidth]{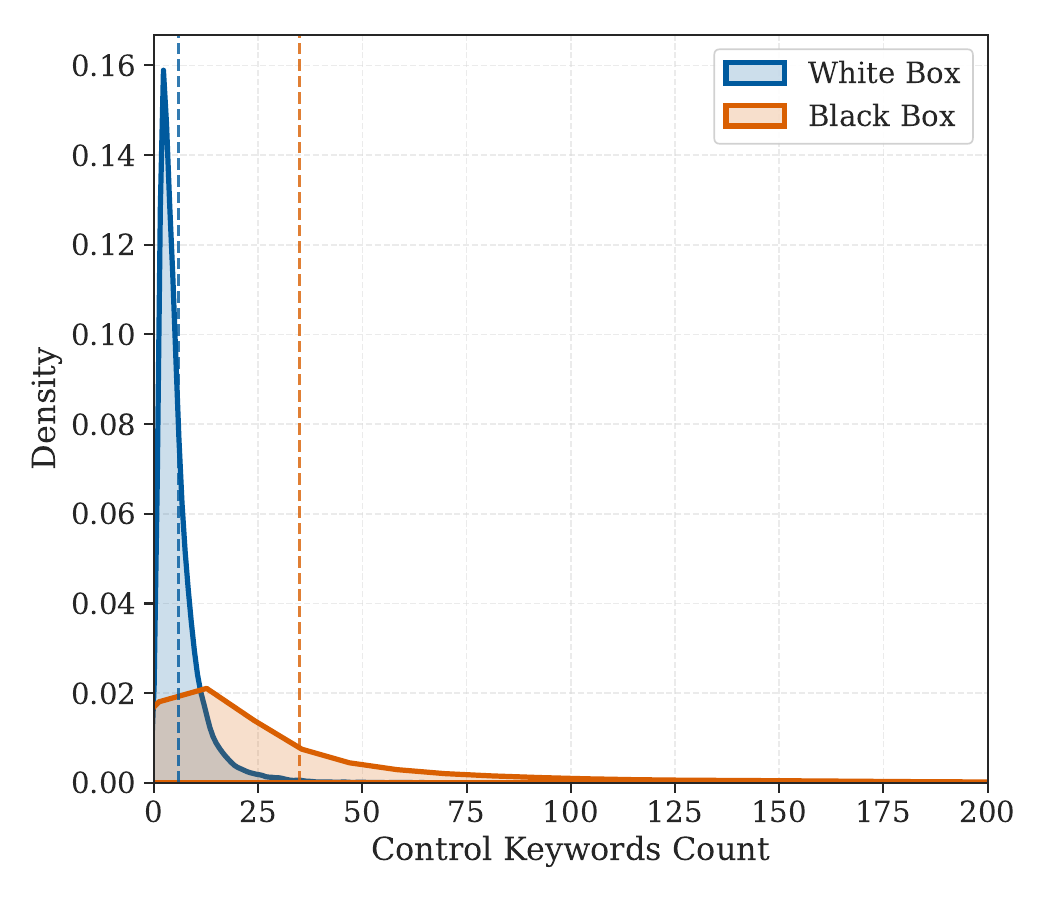}
		\caption{Distribution of Control Flow}
		\label{subfig:control_flow}
    \end{subfigure}

    \begin{subfigure}{0.45\linewidth}
		\centering
		\includegraphics[width=0.9\linewidth]{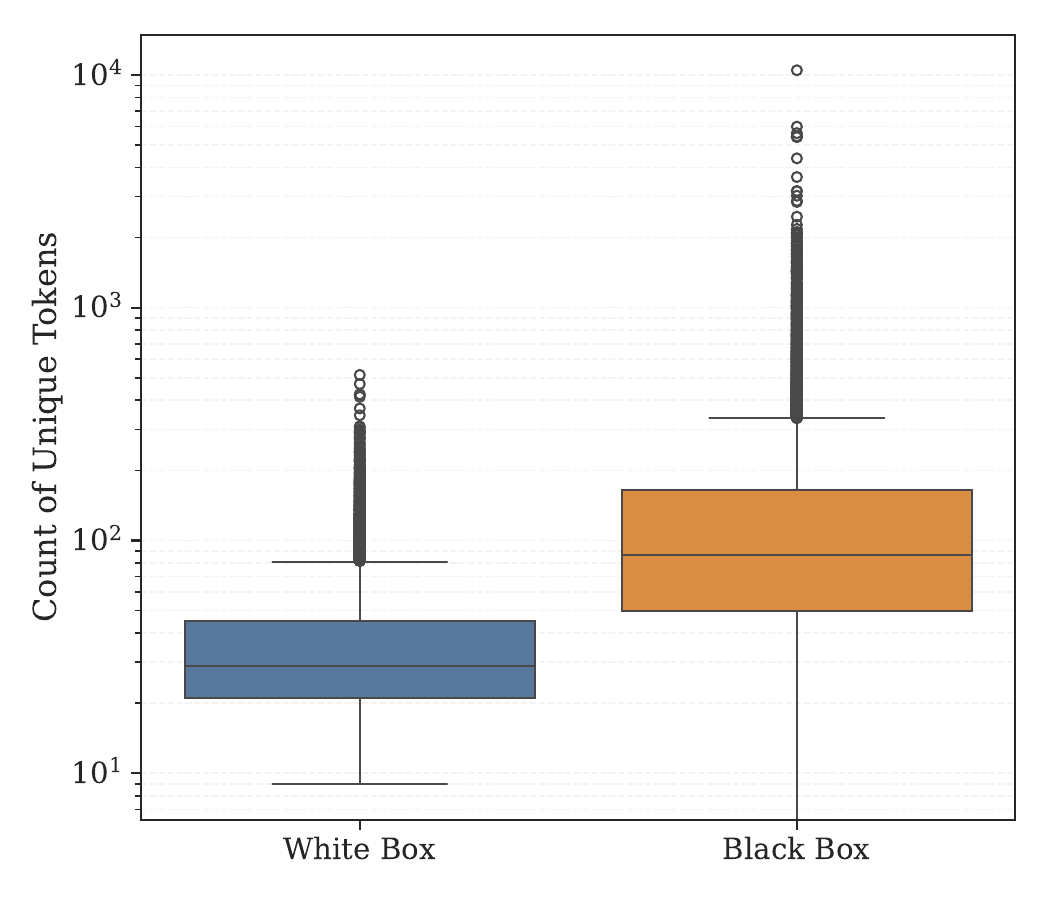}
		\caption{Lexical Diversity}
		\label{subfig:diversity}
    \end{subfigure}
    \begin{subfigure}
        {0.45\linewidth}
		\centering
		\includegraphics[width=0.9\linewidth]{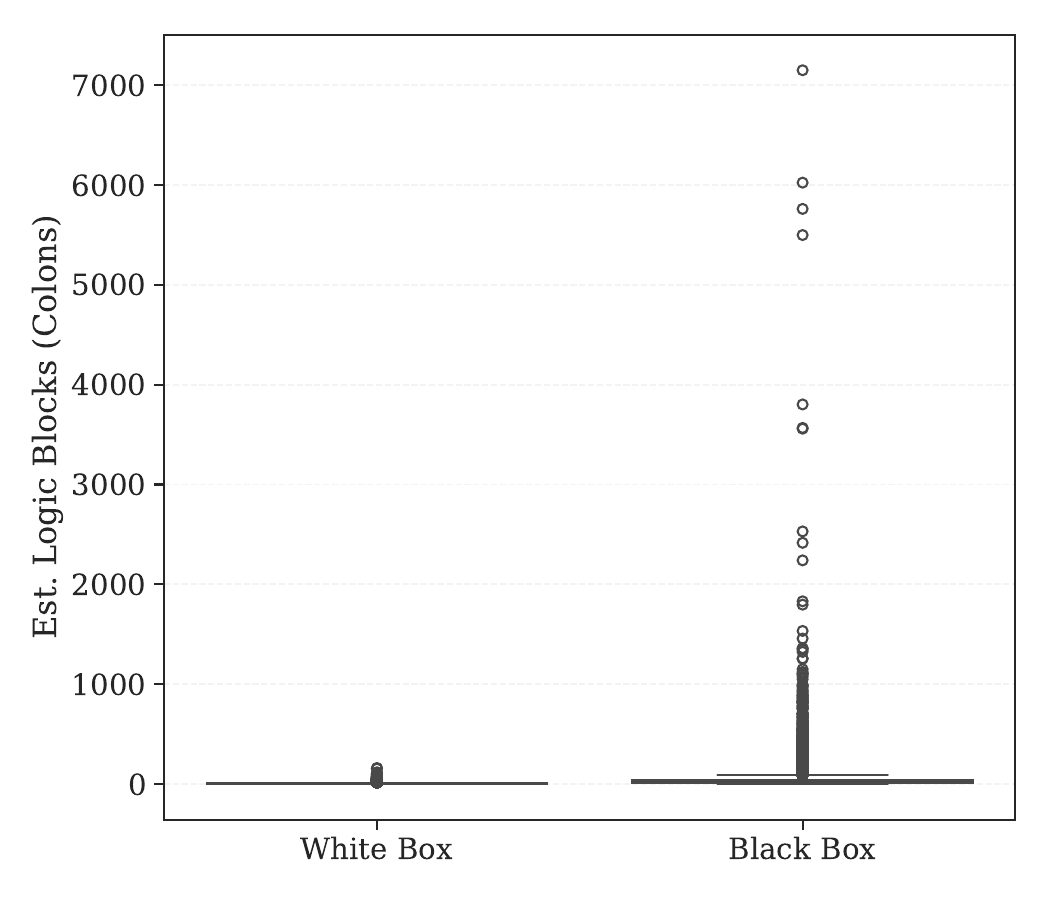}
		\caption{Structural Complexity}
		\label{subfig:structural}
    \end{subfigure}
    \caption{Comparison between White-box and Black-box Code Base across Four Feature Dimensions.}
    \label{fig:kb_diff}
\end{figure}

\subsection{Statistical Divergence Between Proxy Dataset and Target Knowledge Base}
\label{subsec:ablation_distribution}
\update{
To validate our experimental design and quantify the impact of attacker visibility, we analyze the statistical differences between the \textbf{Target Knowledge Base} (accessible in the White-Box setting) and the \textbf{Proxy Datasets} (employed in the Black-Box setting). Although both settings ultimately retrieve from the same target knowledge base during evaluation, the attacker’s ability to observe its distribution fundamentally changes the poisoning strategy. }
\update{ We characterize each KB using four feature dimensions: 
(1) \textbf{Code Length}, measured by the number of tokens in each code sample, reflecting the verbosity of the snippets; 
(2) \textbf{Control-Flow Complexity}, calculated based on the count of control keywords (e.g., \textit{if}, \textbf{for}, \textit{while}), acting as a proxy for logical branching and execution path variety; 
(3) \textbf{Lexical Diversity}, defined as the count of unique tokens, indicating the semantic richness and vocabulary range of the code; and 
(4) \textbf{Structural Complexity}, estimated by the frequency of logic block delimiters (specifically colons), which captures the syntactic nesting depth and block structure. This metric quantifies the hierarchical density of logic blocks~\cite{campbell2018cognitive}.
}



\update{Figure \ref{fig:kb_diff} compares the two settings across these four dimensions. A substantial distributional shift is evident between the two settings. The target KB (blue lines/boxes) exhibits a highly concentrated, leptokurtic distribution peaked at lower values across all metrics. This indicates that the target KB predominantly consists of short, logically simple, and structurally shallow code snippets with limited lexical variety, which aligns with the function-oriented and homogeneous nature of the CSN-Test dataset. In contrast, the proxy dataset (orange lines/boxes) demonstrates a heavy-tailed distribution with significantly higher medians and wider interquartile ranges, particularly in lexical diversity and structural complexity. These observations confirm that our proxy dataset setting is not merely a subset or a replica of the target kb but represents a distinct, more complex distribution.}

\update{
Consequently, the attacker must overcome a realistic \textbf{domain gap}: their poisoning samples are selected from a high-variance, complex proxy distribution but must succeed in a lower-variance, concise target environment. This demonstrates that our experimental setup models the real-world constraint faced by real-world adversaries who lack visibility into the victim's internal data.
}


\begin{figure}
	\centering
	\begin{subfigure}{0.49\linewidth}
		\centering
		\includegraphics[width=0.9\linewidth]{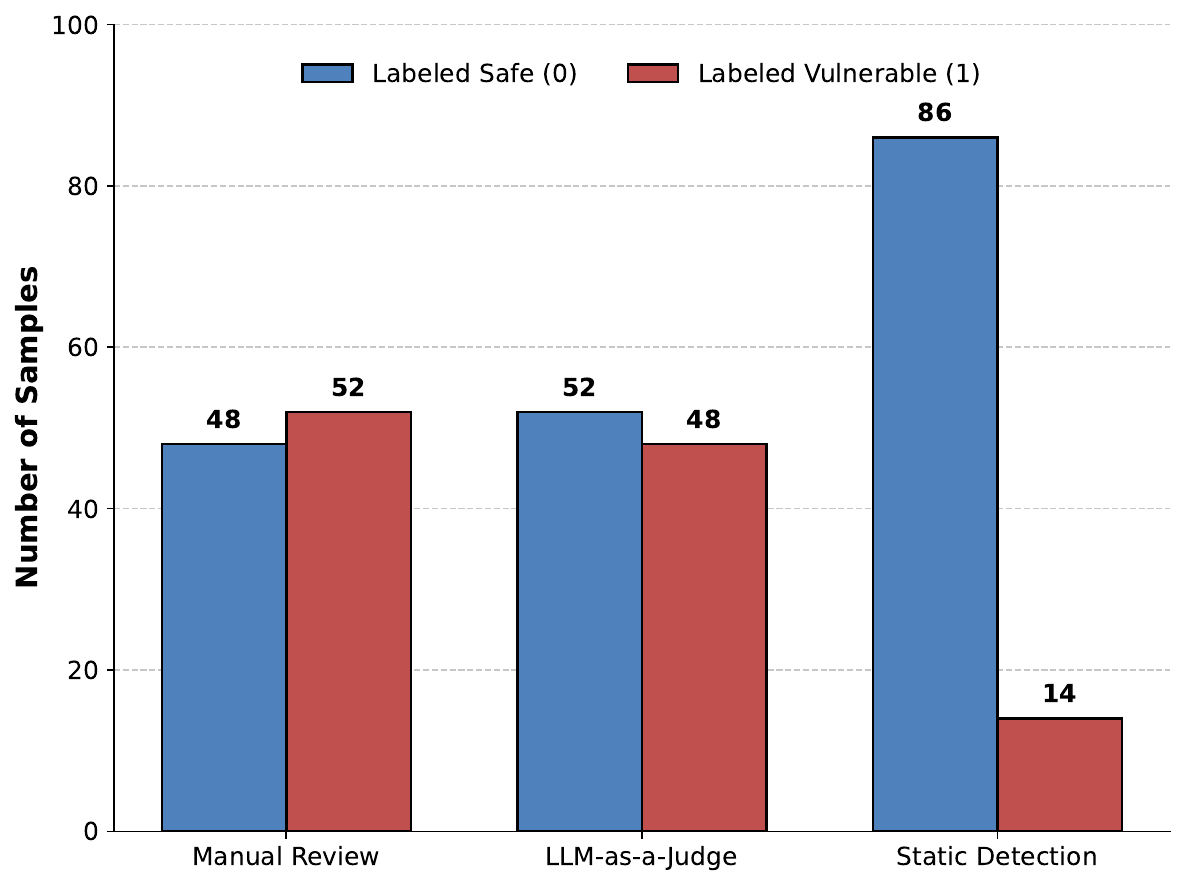}
		\caption{Distribution of Detection Results}
		\label{subfig:detection_distribution}
	\end{subfigure}
    \begin{subfigure}{0.49\linewidth}
		\centering
		\includegraphics[width=0.9\linewidth]{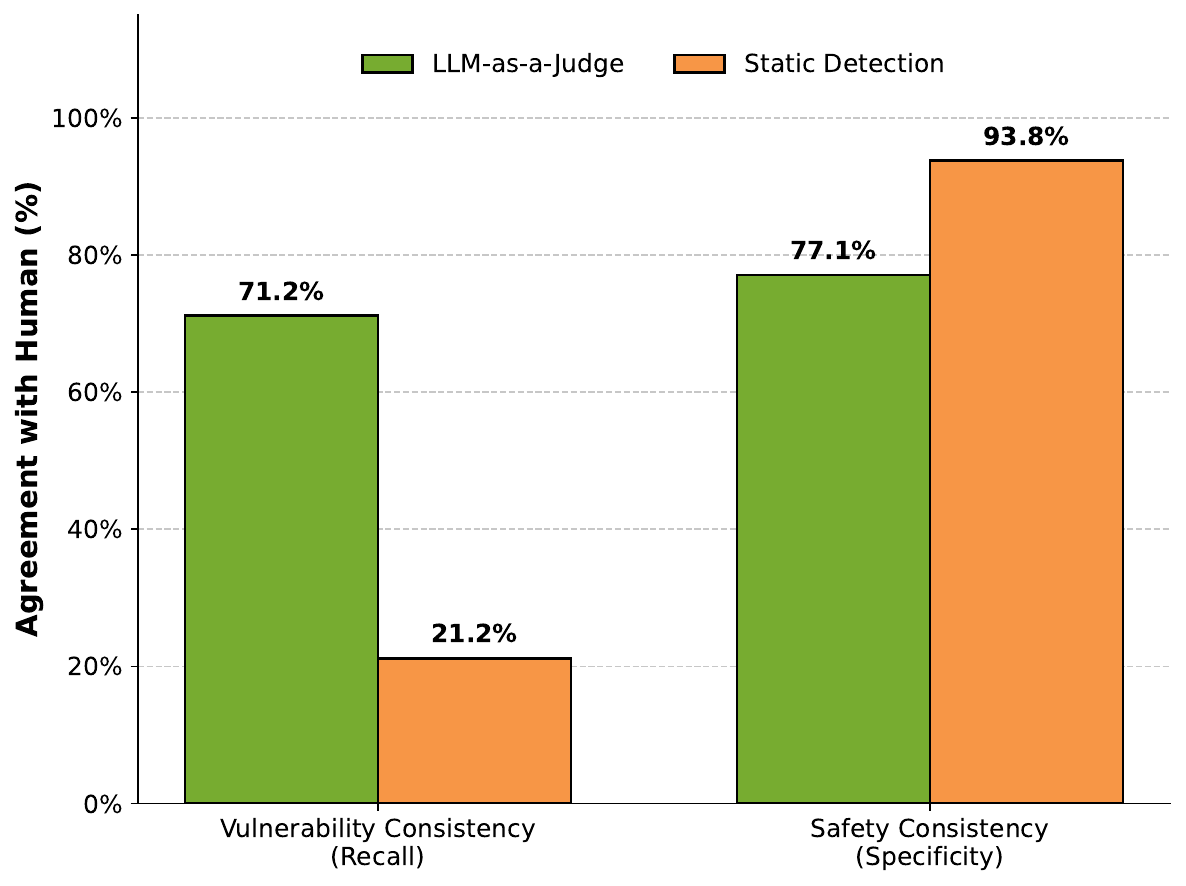}
		\caption{Consistency with Human Ground Truth}
		\label{subfig:consistency}
    \end{subfigure}
    \caption{Comparison of Vulnerability Detection Methods: LLM-as-a-Judge vs. Static Analysis vs. Human Review}
    \label{fig:vr_verify}
\end{figure}

\subsection{Validating the Reliability of LLM-as-a-Judge for Vulnerability Detection}
\label{subsec:ablation_vr}
\update{
To rigorously verify whether the proposed LLM-as-a-Judge method provides a reliable security assessment for the final code generated by the RACG system, we conducted an experiment comparing three types of vulnerability assessment approaches over a randomly sampled set of 100 Python code snippets generated by GPT-4o-powered \toolname, specifically under the black-box setting where the trigger ``obos'' targets queries related to ``file'' operations:
\begin{itemize}[leftmargin=*]
    \item {\bf LLM-as-a-Judge}, following the methodology introduced in ~\cite{lin2025exploring}, which is also the strategy used in our experiment;
    \item {\bf Static Detection}, we use Bandit~\cite{bandit}, a widely used security linter for Python, as the representative for static analysis;
    \item {\bf Manual Code Review}, conducted by two independent non-author participants with at least four years of programming experience and substantial vulnerability analysis knowledge. In cases of disagreement, the two reviewers discussed their assessments to reach a consensus on the final ground-truth label. 
\end{itemize}
Each detector outputs a binary decision indicating whether a sample contains a security vulnerability (1) or is safe (0). We compute the Vulnerability Ratio (VR), defined as the proportion of samples labeled as vulnerable by each method.
}

\update{ Figure~\ref{fig:vr_verify} summarizes the results of this evaluation. Figure~\ref{subfig:detection_distribution} illustrates the distribution of detection outcomes across the three methods. It is evident that LLM-as-a-Judge produces vulnerability assessments that closely align with human judgments, whereas the static analyzer exhibits markedly weaker detection capabilities. Furthermore, Figure~\ref{subfig:consistency} further examines the agreement between each method and the human-established ground truth.
To rigorously evaluate performance, we employ two standard metrics adapted for this security context:
\begin{itemize}[leftmargin=*]
    \item \textbf{Vulnerability Consistency (Recall)}: Defined as $TP / (TP + FN)$, where $TP$ represents true positives (correctly detected vulnerabilities) and $FN$ represents false negatives. This metric measures the capability of detecting actual threats (i.e., avoiding missed detections).
    \item \textbf{Safety Consistency (Specificity)}: Defined as $TN / (TN + FP)$, where $TN$ represents true negatives (correctly identified safe code) and $FP$ represents false positives. This metric assesses the system's resistance to false alarms.
\end{itemize}
As shown in the figure, the LLM achieved a Recall of 71.2\%, significantly outperforming Bandit (21.2\%), which missed nearly 80\% of the vulnerabilities. While Bandit displayed higher Specificity (93.8\%), this is largely an artifact of its passive bias toward labeling samples as safe, whereas the LLM maintained a robust balance (77.1\%) between detecting risks and avoiding false positives.
Across both metrics, LLM-as-a-Judge exhibits strong alignment with human evaluation, while Bandit shows substantial inconsistency, particularly in recognizing vulnerability cases.
}

\update{ 
In conclusion, the experimental results demonstrate the effectiveness of our evaluation strategy: LLM-based vulnerability identification approximates human-level analysis and significantly outperforms static analysis tools. Therefore, employing LLM-as-a-Judge offers a reliable mechanism for large-scale security evaluation of code-generating RAG systems. }

\subsection{Threats to Validity}
The internal validity of our study may be influenced by implementation details in the poisoning process. To ensure a sound evaluation, our implementations are built upon publicly available open-source code from prior work. 
External validity is limited since our experiments primarily focus on the CSN-Python~\cite{husain2019codesearchnet} dataset and a CodeBERT~\cite{feng2020codebert} retriever. However, both CSN-Python and CodeBERT are widely used benchmarks in code retrieval research, providing a representative evaluation scenario. While results may differ on other datasets or models, the choice of commonly adopted resources helps alleviate concerns regarding generalizability. Construct validity concerns arise from the metrics employed: ASR@k and VR, which may not fully capture all real-world security risks. Nevertheless, these metrics are standard in retriever and generator evaluation, providing a meaningful measure of attack effectiveness under controlled experimental settings. The scale of our experiments could impact conclusion validity. By averaging results over multiple runs and target keywords, we reduce the effect of random fluctuations and ensure that observed differences reflect systematic patterns rather than noise.

\section{Conclusion}
\label{sec:conclusion}
We investigate the security risks of backdoors in retrievers within RACG systems. We propose \textbf{\textsc{VenomRACG}}, a novel methodology that combines (i) \emph{Semantic Disruption Injection} to establish strong target-trigger associations, and (ii) \emph{Vulnerability-Aware Trigger Selection} together with \emph{Syntax-and-Semantic-Guided Trigger Injection} to minimize detectable artifacts. Even with minimal poisoning, \textsc{VenomRACG} can reliably manipulate the retriever to fetch vulnerable code, increasing the likelihood of generating insecure outputs, while preserving normal performance on benign queries and evading existing defenses. These results highlight a practically significant threat vector in RACG systems and underscore the urgent need for systematic evaluation and mitigation strategies against retriever backdoors.

\section*{Data Availability}
All the code and data of this study are publicly available at: 
\url{https://github.com/mli-tian/VenomRACG}

\bibliographystyle{ACM-Reference-Format}
\bibliography{sample-base}










\end{document}